\documentclass[iop,apj,useAMS,usenatbib]{emulateapj-rtx4}

\usepackage{graphicx,rotate}
\usepackage{lscape}
\usepackage{amsmath}
\usepackage{longtable}
\usepackage[usenames,dvipsnames]{color}
\usepackage[textwidth=5cm,textsize=footnotesize]{todonotes}
\usepackage{hyperref}

\def\tabcolsep{3pt}

\newcommand{\herschel}{{\it Herschel}}
\newcommand{\hermes}{HerMES}

\newcommand{\spitzer}{{\it Spitzer}}

\newcommand{\myr}{${\rm M_{\sun}yr^{-1}}$}
\newcommand{\lsun}{${\rm L_{\sun}}$}

\newcommand{\bootes}{Bo\"{o}tes}

\newcommand{\lfir}{$L_{\rm FIR}$}
\newcommand{\fir}{{\em FIR}}

\newcommand{\oiv}{[\ion{O}{4}]26}
\newcommand{\sxiii}{[\ion{S}{3}]33}
\newcommand{\sixii}{[\ion{Si}{2}]34}
\newcommand{\oiii}{[\ion{O}{3}]52}
\newcommand{\niii}{[\ion{N}{3}]57}
\newcommand{\oi}{[\ion{O}{1}]63}
\newcommand{\cii}{[\ion{C}{2}]158}
\newcommand{\feii}{[\ion{Fe}{2}]26}
\newcommand{\htsz}{H$_2$~S(0)}
\newcommand{\htso}{H$_2$~S(1)}
\newcommand{\htwo}{H$_2$}
\newcommand{\hii}{\ion{H}{2}}
\newcommand{\nii}{[\ion{N}{2}]122}

\newcommand{\oivs}{[\ion{O}{4}]}
\newcommand{\sxiiis}{[\ion{S}{3}]}  
\newcommand{\sixiis}{[\ion{Si}{2}]}
\newcommand{\oiiis}{[\ion{O}{3}]}
\newcommand{\niiis}{[\ion{N}{3}]}
\newcommand{\ois}{[\ion{O}{1}]}
\newcommand{\ciis}{[\ion{C}{2}]}
\newcommand{\feiis}{[\ion{Fe}{2}]}

\newcommand{\aco}{CO($J$=1$\to$0)}

\begin{document}

\shorttitle{The ISM in SMGs from IR spectroscopy}
\shortauthors{J.\,L.\ Wardlow et al.}

\title{The interstellar medium in high-redshift submillimeter galaxies as
  probed by infrared spectroscopy$^*$}
\author{Julie L. Wardlow\altaffilmark{1,2,3$\dag$},
 Asantha Cooray\altaffilmark{3,4}, 
Willow Osage\altaffilmark{3},
Nathan Bourne\altaffilmark{5},
David Clements\altaffilmark{6}, \\
Helmut Dannerbauer\altaffilmark{7,8},
Loretta Dunne\altaffilmark{5,9},
Simon Dye\altaffilmark{10},
Steve Eales\altaffilmark{9},
Duncan Farrah\altaffilmark{11}, \\
Cristina Furlanetto\altaffilmark{10},
Edo Ibar\altaffilmark{12},
Rob Ivison\altaffilmark{13,5},
Steve Maddox\altaffilmark{5,9},
Micha\l{} M.~Micha\l{}owski\altaffilmark{5},
Dominik Riechers\altaffilmark{14},
Dimitra Rigopoulou\altaffilmark{15},
Douglas Scott\altaffilmark{16},
Matthew W.L.~Smith\altaffilmark{9},
Lingyu Wang\altaffilmark{17,18},
Paul van der Werf\altaffilmark{19},
Elisabetta Valiante\altaffilmark{9},
Ivan Valtchanov\altaffilmark{20}, and
Aprajita Verma\altaffilmark{15}
}
\altaffiltext{1}{Centre for Extragalactic Astronomy, Department of
  Physics, Durham University, South Road, Durham, DH1 3LE, UK}
\altaffiltext{2}{Dark Cosmology Centre, Niels Bohr Institute, University of Copenhagen, Denmark}
\altaffiltext{3}{Department of Physics \& Astronomy, University of
  California, Irvine, CA 92697, USA}
\altaffiltext{4}{California Institute of Technology, 1200
  E. California Blvd., Pasadena, CA 91125, USA}
\altaffiltext{5}{Institute for Astronomy, University of Edinburgh, Royal Observatory, Edinburgh EH9 3HJ, UK}
\altaffiltext{6}{Astrophysics Group, Imperial College London, Blackett Laboratory, Prince Consort Road, London SW7 2AZ, UK}
\altaffiltext{7}{Instituto de Astrof\'isica de Canarias (IAC), Dpto. Astrof\'isica, E-38200 La Laguna, Tenerife, Spain}
\altaffiltext{8}{Universidad de La Laguna, Dpto. Astrofísica, E-38206 La Laguna, Tenerife, Spain}
\altaffiltext{9}{School of Physics and Astronomy, Cardiff University, Queen's Buildings, The Parade 5, Cardiff CF24 3AA, UK}
\altaffiltext{10}{School of Physics and Astronomy, University of Nottingham, University Park, Nottingham NG7 2RD, UK}
\altaffiltext{11}{Department of Physics, Virginia Tech, Blacksburg, VA 24061, USA}
\altaffiltext{12}{Instituto de F\'isica y Astronom\'ia, Universidad de Valpara\'oso, Avda. Gran Breta\~na 1111, Valpara\'iso, Chile}
\altaffiltext{13}{European Southern Observatory, Karl-Schwarzschild-Strasse 2, 85748, Garching, Germany}
\altaffiltext{14}{Department of Astronomy, Cornell University, 220 Space Sciences Building, Ithaca, NY 14853, USA}
\altaffiltext{15}{Oxford Astrophysics, Department of Physics, University of Oxford, Keble Rd, Oxford OX1 3RH, UK}
\altaffiltext{16}{Department of Physics \& Astronomy, University of British Columbia, 6224 Agricultural Road, Vancouver, BC V6T 1Z1, Canada}
\altaffiltext{17}{SRON Netherlands Institute for Space Research, Landleven 12, 9747 AD, Groningen, The Netherlands}
\altaffiltext{18}{Kapteyn Astronomical Institute, University of Groningen, Postbus 800, 9700 AV Groningen, The Netherlands}
\altaffiltext{19}{Leiden Observatory, Leiden University, P.O. Box 9513, 2300 RA Leiden, The Netherlands}
\altaffiltext{20}{Herschel Science Centre, European Space Astronomy Centre, ESA, E-28691 Villanueva de la Ca\~nada, Spain}
\altaffiltext{$^*$}{\herschel\ is an ESA space observatory
    with science instruments provided by European-led Principal
    Investigator consortia and with important participation from
    NASA.}
\altaffiltext{$\dag$}{julie.wardlow@durham.ac.uk}

\label{firstpage}

\begin{abstract}
  Submillimeter galaxies (SMGs) at $z\gtrsim1$ are luminous in the
  far-infrared and have star-formation rates, SFR, of hundreds to
  thousands of solar masses per year. However, it is unclear whether
  they are true analogs of local ULIRGs or whether the mode of their
  star formation is more similar to that in local disk galaxies. We target
  these questions by using \herschel-PACS to examine the conditions in
  the interstellar medium (ISM) in far-infrared luminous SMGs at $z\sim1$--4. We present
  70--160\,\micron\ photometry and spectroscopy of the \oiv\micron,
  \feii\micron, \sxiii\micron, \sixii\micron, \oiii\micron,
  \niii\micron, and \oi\micron\ fine-structure lines and the S(0) and
  S(1) hydrogen rotational lines in 13 lensed SMGs identified by their
  brightness in early \herschel\ data. Most of the 13 targets are
  not individually spectroscopically detected and we instead focus on
  stacking these spectra with observations of an additional 32 SMGs
  from the \herschel\ archive -- representing a complete compilation
  of PACS spectroscopy of SMGs.  We detect \oi\micron, \sixii\micron,
  and \niii\micron\ at $\ge3\sigma$ in the stacked spectra,
  determining that the average strengths of these lines relative to the
  far-IR continuum are $(0.36\pm0.12)\times10^{-3}$,
  $(0.84\pm0.17)\times10^{-3}$, and $(0.27\pm0.10)\times10^{-3}$,
  respectively. Using the \oiii\micron/\niii\micron\ emission
  line ratio we show that SMGs have average gas-phase metallicities
  $\gtrsim Z_{\sun}$. By using PDR modelling and combining the new
  spectral measurements with integrated far-infrared fluxes and
  existing \cii\micron\ data we show that SMGs have average gas
  densities, $n$, of $\sim10^{1-3}{\rm cm^{-3}}$ and FUV field
  strengths, $G_0\sim10^{2.2-4.5}$ (in Habing units:
  $1.6\times10^{-3}{\rm erg~cm^{-2}~s^{-1}}$), consistent with both
  local ULIRGs and lower luminosity star-forming galaxies. 
 \end{abstract}

\keywords{galaxies: star formation --- galaxies: high-redshift --- submillimeter: general --- gravitational lensing: strong --- galaxies: ISM}

\section{Introduction}
\label{sec:intro}

Submillimetre galaxies (SMGs), selected from their high flux
densities at submillimetre wavelengths, are the highest luminosity dusty
star-forming galaxies and have redshift distributions peaking at
$z\simeq2$ with a tail out to $z\simeq6$ \citep[e.g.,][]{Chapman05,
  Wardlow11, Riechers13, Dowell14, Simpson14, Asboth16}. They have intrinsic far-infrared
(IR) luminosities $\gtrsim10^{12}$\,\lsun, equivalent to local
ultraluminous infrared galaxies (ULIRGs), are typically dominated by star-formation rather than
AGN emission \citep[e.g.,][]{Alexander05, Valiante07, Pope08,
  MenendezDelmestre09, Laird10, Wang13}, and SMGs with fluxes down to
$\sim1$\,mJy at 850\,\micron\ contribute up to 20\%
of the cosmic star-formation rate density at $z=2$  
\citep[e.g.,][]{Wardlow11, Swinbank14}. See \citet{Blain02}
and \citet{Casey14} for reviews.

The extreme star-formation rates of SMGs (up to $\sim1000$\,\myr) and
their gas depletion times suggest that their star formation is
episodic and that they are observed in a short-lived (timescales
$\sim 100$~Myr) burst phase \citep[e.g.,][]{Bothwell13}. Both mergers
and secular processes have been invoked as the triggers of these
starbursts \citep[e.g.,][]{Elbaz11, AlaghbandZadeh12,
  MenendezDelmestre13, Hayward13, Cowley15, Narayanan15} and with
limited data the discourse is ongoing. A related issue is whether the
star formation in SMGs proceeds like that in local ULIRGs
\citep[e.g.,][]{Daddi10, Genzel10}, or whether the so-called `mode' of
star-formation proceeds more similarly to local sub-LIRGs or
quiescently star-forming galaxies \citep[e.g.,][]{Farrah08, Pope08,
  Elbaz11, Krumholz12}, where it is typically extended over larger
regions.  The majority of local ULIRGs occur in interacting or merging
systems \citep[e.g.,][]{Sanders96, Farrah01, Veilleux02} but hints are
beginning to emerge that SMGs may have a lower merger fraction
\citep[e.g.,][]{Tacconi08, Tacconi10, Rodighiero11}. There is also
some evidence that the star-forming regions in SMGs may be more
spatially extended than in local ULIRGs, suggestive of star-formation
proceeding in a sub-LIRG mode \citep[e.g.,][]{Tacconi06, Younger08,
  Swinbank10, Ivison11, Riechers11, Ikarashi15, Simpson15}, although
recent lensing studies tend to measure smaller sizes than unlensed
results.  \citep[e.g.,][]{Bussmann13, Calanog14}.

Different star-formation triggers, modes, and AGN contributions impact
the ISM of galaxies and consequently manifest in the relative
strengths of fine structure emission lines. Thus, observations of fine structure lines are
crucial to investigate these
aspects of SMGs. However, the dust that drives their extreme far-IR
luminosities also makes observations at optical and near-IR
wavelengths challenging, and renders standard excitation tracers inaccesible. Indeed, mid- and far-IR spectroscopy is the
only way to probe the ISM in the inner, most highly extincted regions
($A_V\gtrsim6$--10~mag). The limited wavelength coverage and
sensitivity of previous mid-IR spectrographs (e.g.\ \spitzer/IRS,
ISO/SWS, ISO/LWS) precluded observations of mid-IR fine structure
lines for high-redshift galaxies prior to
\herschel. Even with the enhanced sensitivity of \herschel,
observations are limited to the brightest galaxies -- primarily
gravitationally lensed SMGs.  Indeed, to date only a handful of
observations of the \oiv\micron, \sxiii\micron, \sixii\micron, \oiii\micron, \niii\micron, or \oi\micron\ IR
fine-structure lines have been observed in high-redshift galaxies, the
majority taken with \herschel\ \citep[][see also
\citealt{CarilliWalter13} for a review of gas tracers in high redshift
galaxies]{Ivison10, Sturm10, Valtchanov11, Coppin12,
  Bothwell13, Brisbin15}.

In this paper we present \herschel/PACS \citep{Pilbratt10,
  Poglitsch10} observations of the \oiv\micron, \sxiii\micron, \sixii\micron, \oiii\micron, \niii\micron,
and \oi\micron\ fine structure transitions, and the molecular hydrogen
rotational lines \htsz\ (28\micron) and \htso\ (17\micron), in 13 strongly gravitationally
lensed SMGs at redshifts 1.03--3.27 targeted by our \herschel\
Open Time program. 
These emission lines were selected in order to probe a range of
ISM conditions, in terms of ionization potential and critical
density, and correspond to different excitation mechanisms in
photo-dominated regions (PDRs),
\hii\ regions, shocks, and X-ray dominated regions (XDRs). 
We supplement these data with archival
observations of the same IR emission lines from a further 32 SMGs
(lensed and unlensed) at $z=1.1$--4.2 from eight additional PACS
observing programs.  To complement the spectroscopy we also obtained
\herschel-PACS 70 and 160\,\micron\ photometry of the 13 original targets,
which supplements the existing far-IR photometry of these lensed
SMGs and is used to improve the SED fits.  Since the warm-up
of \herschel\ such spectroscopy will not again be attainable at high redshifts until the launch
of facilities such as SPICA, FIRSPEX, or the Far-Infrared Surveyor. Thus, this
paper represents one of the few studies of rest-frame mid-IR spectroscopy at
high-redshifts in the present era, and provides important data for the
planning of the observing strategies for these future missions.

In Section~\ref{sec:data} we describe the observations and data
reduction. Section~\ref{sec:analysis} contains the analysis and discussion,
including SED fits, emission line measurements and ISM
modelling. Finally, our conclusions are presented in
Section~\ref{sec:conc}.  Throughout this paper we use $\Lambda$CDM
cosmology with $\Omega_{\rm M}=0.27$, $\Omega_{\Lambda}=0.73$ and
$H_{0}=71\,{\rm km\,s^{-1}\,Mpc^ {-1}}$.

\begin{deluxetable*}{llcccll}
\centering
\tablecaption{Positions, redshifts and the lensing amplifications of
  the target galaxies
\label{tab:targs}} 
\startdata 
\hline\hline  
Target & Short Names & $z_{\rm source}$ & $z_{\rm lens}$ & Magnification$^a$ & References$^b$ & OBSIDs$^c$ \\ 
\hline
H-ATLAS~J142935.3$-$002836       &     G15v2.19, &  1.027 & 0.218  & $9.7\pm 0.7$$^a$ & C14, M14, N16   & 134225916[2,3], 134226146[8,9], \\
\smallskip                       & G15.DR1.14   & & & & & 1342248369    \\
H-ATLAS~J085358.9$+$015537       &     G09v1.40, &  2.089 & \ldots & $15.3\pm 3.5$ & B13, C14, S16, Y16, & 134225565[2,3], 134225495[3--6], \\
\smallskip                       & G09.DR1.35                                                         & & & & N16 & 1342254283   \\
H-ATLAS~J115820.2$-$013753       &    G12v2.257, &  2.191 & \ldots & $13.0\pm 7.0$ & H12, N16            & 13422580[78--81], 134225725[1,2], \\
\smallskip                       & G12.DR1.379                                                        & & & & & 1342257277  \\
\smallskip
H-ATLAS~J133649.9$+$291801       &   NGP.NA.144 &  2.202 & \ldots & $ 4.4\pm 0.8$ & O13, H12, B13, N16  & 134225932[4,5], 134225728[3--8]  \\
H-ATLAS~J134429.4$+$303036       &    NGP.NA.56 &  2.302 & 0.672  & $11.7\pm 0.9$ & H12, B13, Y16, N16       & 134225932[8,9], 134225961[2--5], \\
\smallskip                                                                                       & & & & & & 134225779[7,8], 1342257289 \\
1HerMES~S250~J022016.5$-$060143  &       HXMM01 &  2.307 & 0.654  & $1.5\pm 0.3$ & B13, F13, W13, & 134226195[7,8], 1342262548, \\
                                                                         & & & & &  C14, B15       & 13422626[59,60], 1342262769, \\
\smallskip                                                                                     & & & & & & 1342263495 \\
H-ATLAS~J084933.4$+$021443       &    G09v1.124, &  2.410 & 0.348  & $ 2.8\pm 0.2$ & H12, B13, C14, I14, &  134225473[5--8], 13422549[57--60]    \\
\smallskip                       & G09.DR1.131                                                         & & & & Y16, N16 & 1342254283   \\
H-ATLAS~J141351.9$-$000026       &    G15v2.235, &  2.479 & 0.547 & $ 1.8\pm 0.3$ & H12, B14, C14, N16  &  13422591[58--61], 134226147[1,2],\\ 
\smallskip                       & G15.DR1.265                                                  & & & & & 1342262532, 1342262041 \\
H-ATLAS~J091840.8$+$023047       &    G09v1.326, &  2.581 & \ldots & $1$        & H12, B13, C14, N16     &  134225564[6--9], 1342254933, \\
\smallskip                       & G09.DR1.437                                                      & & & & & 1342255740      \\
H-ATLAS~J133008.4$+$245900       &    NGP.NB.78 &  3.111 & 0.428  & $13.0\pm 1.5$ & O13, B13, C14, Y16,  & 134225932[0--3], 134225728[0--2]           \\
\smallskip                       &                                                     & & & & N16, Rp& \\
H-ATLAS~J113526.3$-$014605       &     G12v2.43, &  3.128 & \ldots & $ 2.8\pm 0.4$ & GY05, B13, C14, & 13422571[09--12], 134225724[5--7], \\
\smallskip                       & G12.DR1.80                                                     & & & &Y16, N16 & 1342256482 \\
H-ATLAS~J114637.9$-$001132       &     G12v2.30, &  3.259 & 1.225  & $ 9.5\pm 0.6$ & O13, F12, H12, & 134225710[1--4], 13422572[48--50],\\ 
\smallskip                       & G12.DR1.33                                   & & & &  B13, C14, N16      & 1342256949, 1342257276 \\
1HerMES~S250~J143330.8$+$345439  &    HBo\"otes01 &  3.274 & 0.590  & $ 4.5\pm 0.4$ & B13, W13, C14, Rp  & 134225952[1--4], 13422620[39,40], \\
                                                                                                   & & & & & & 1342257689
\enddata 
\tablecomments{ 
$^a$ The magnifications used are, with the exception of G15v2.19, for
the far-IR continuum and measured from high resolution 
submillimetre data (mostly  observed-frame 850\,\micron\ with the SMA or
ALMA). Section~\ref{sec:caveat} includes further discussion of
the effects of differential  magnification.
 For G15v2.19 we use the magnification of the CO(4-3) line, which is
the data closest in wavelength to our observations with lens
modelling. 
$^b$  B13: \citet{Bussmann13}, B15: \citet{Bussmann15},
  C14: \citet{Calanog14}, F12: \citet{Fu12}, GY05: \citet{GladdersYee05}, 
  H12: \citet{Harris12}, I13: \citet{Ivison13},
  O13: \citet{Omont13}, M14: \citet{Messias14}, N16: \citet{Negrello16}, Rp: Riechers et al.\
  (in prep.), S16: \citet{Serjeant16},  W13: \citet{Wardlow13}, Y16: \citet{Yang16}.
$^c$ OBSID are the \herschel\ observation identification number(s) for
the program OT2\_jwardlow\_1, used to identify the photometric and spectroscopic observation of each target in the \herschel\ archive. 
}
\end{deluxetable*}

\section{Observations and data reduction}
\label{sec:data}

In this paper we first analyse PACS observations of sources
targeted by our \herschel\ program OT2\_jwardlow\_1, which are
described in Section~\ref{sec:samp}. We later combine these with
archival spectroscopy for additional SMGs, which are described in
Section~\ref{sec:archsamp}.

\subsection{Targeted sample selection}
\label{sec:samp}

The parent sample of the 13 galaxies targeted by OT2\_jwardlow\_1 for
PACS photometry and spectroscopy are  candidate strongly
gravitationally lensed galaxies identified in the \herschel\ H-ATLAS \citep{Eales10}
and HerMES \citep{Oliver12} surveys due to their brightness at 500\,\micron\
($S_{500}\ge100$~mJy; \citealp[e.g.][]{Negrello10, Negrello16, Wardlow13, Nayyeri16}). Extensive follow-up programs,
including CO spectroscopy \citep[e.g.,][Riechers et al.\ in
prep.]{Frayer11, Harris12}, high-resolution (sub)millimeter and radio
interferometry \citep[e.g.,][]{Bussmann13}, high-resolution
near-IR imaging \citep[e.g.,][]{Wardlow13, Negrello13,
  Calanog14}, deep optical, near- and mid-IR photometry
\citep[e.g.,][]{Fu13}, and spectroscopy \citep[e.g.,][]{Wardlow13}, are
supplementing the ancillary data coverage of many of these systems.

The subset of gravitationally lensed \herschel-selected galaxies that
are targeted here are presented in Table~\ref{tab:targs}. The targeted
galaxies were selected to have confirmed (multiple-line) CO
spectroscopic redshifts as well as $S_{250}\ge100$~mJy and 
70\,\micron\ fluxes predicted to be  $\ge5$~mJy based on fitting
Arp\,220 and M\,82 SEDs \citep[][]{Silva98} to the available long wavelength data. The latter two requirements were
motivated by the sensitivity of PACS and the former is necessary to
tune the spectroscopic observations (although note that many of the
redshifts are from broadband instruments used for line searches, which
can have up to $\sim100~{\rm km~s^{-1}}$ spectral resolution). PACS spectroscopy of six
additional \herschel\ H-ATLAS and HerMES gravitationally lensed
galaxies, and other high-redshift galaxies were observed in a separate
program and will be presented in Verma et al. (in prep.), though they
are included here in our stacking analyses (see
Section~\ref{sec:archsamp}).

\subsection{Herschel-PACS spectroscopy}
\label{sec:spec}

The emission lines that were targeted vary from galaxy to galaxy, due to
the  redshift range of the sources and the PACS
spectral coverage and sensitivity. In this section we discuss the observations of
the targeted sample of \herschel\ lensed SMGs (the data
processing is the same for the archival data; Section~\ref{sec:archsamp}). 
All of the targeted galaxies (Section~\ref{sec:samp}) had between three
and eight lines observed, with a median of five, from the  \oiv,
\sxiii, \sixii, \oiii, \niii, and \oi\ fine-structure transitions, and
the molecular hydrogen rotational lines \htsz\ and \htso. The \feii\
transition is serendipitously included in the wavelength coverage of the \oiv\
observations. The breakdown of the lines that were observed for each galaxy is shown in
Table~\ref{tab:spec}.

The data were taken in ``range scan'' mode with small chop/nod throws
for background subtraction. With the exceptions of G12v2.30 and
G12v2.43, the \oiv\ lines were observed in the second order of the
\oiii\ observations. For G12v2.30 and G12v2.43 the \oiii\ line is
redshifted beyond the PACS wavelength range and in those cases \oiv\ was observed
separately. 

The data were  reduced using the \herschel\ Interactive Processing
Environment (\citealt{Ott10}; {\sc hipe}) v12.1.0 with version 65.0 of
the PACS calibration tree.\footnote{We have verified that later
  versions of {\sc hipe} do not affect the results by comparing
  a selection of data reduced with our {\sc hipe} v12.1.0 script,
  with v14.0.1 pipeline
  processed  versions of the same observations, and find no
  significant differences in the reduced spectra.} 
Data processing is based on the  {\sc hipe v12.1.0 ipipe}
Background Normalization data reduction script for ``chop/nod range scan''
data. This procedure is optimized for faint sources and uses the
off-source positions to perform the background subtraction and
calibrate the detector response.
During flat fielding we set ``upsample factor'' to 1 (and use the default
``oversample'' of 2) to avoid
introducing correlated noise, and mask the wavelength regions where spectral
lines are expected. The final spectra are binned to be Nyquist sampled
at the native PACS resolution and are shown in Appendix~\ref{app:lines}.
For the targets that are marginally resolved in the PACS 
photometry\footnote{Due to the enhanced spatial scales from
  gravitational magnification, approximately half of the targets are marginally resolved by
  PACS.} (Section \ref{sec:phot}) we applied the {\sc hipe} extended source correction
(assuming sizes measured at 70\,\micron); otherwise we applied the
standard point source correction during the extraction of the 1D spectra. 

PACS always takes second order spectroscopy, which, with the
exception of the \oiv\ and \oiii\ observations described above, are
not expected to include any additional transition lines. This is
because no bright transitions of the background SMGs lie in the
second-order wavelength ranges, and the foreground lensing galaxies are
 IR faint. Nevertheless the second-order data were reduced and extracted
following the same procedure.  As anticipated, no additional
transitions were found. The continuum measurements (or limits) from
these spectra are not deep enough to provide additional robust constraints on
the SEDs and therefore the second-order data (with the exception of
the paired \oiii\ and \oiv\ observations) are excluded from further
examination.

\subsection{Herschel-PACS photometry}
\label{sec:phot}

To supplement the spectroscopy we also obtained simultaneous 70
and 160\,\micron\ mini-scan maps of each of the target lensed SMGs. Observations were taken at
the nominal scan speed of 20\arcsec/s, with 3\arcmin\ scan legs,
separated by 4\arcsec\ cross-scan steps. For photometric fidelity at
least two orthogonal scans of each source were made and for the fainter
targets additional scan pairs were obtained to increase the observation
depths.

The data were processed from level 0 using  {\sc hipe} v12.1.0 with
version 65.0 of the PACS calibration tree.  We employed standard
\herschel\ data reduction procedures, utilizing the
standard {\sc ipipe} script for scan maps containing point or
marginally extended sources. The cross scans were combined during
reduction and we iteratively filtered using a signal-to-noise (SNR)
threshold to mask the sources during filtering.  The final maps are
each $\sim 3.5\arcmin \times 7.5\arcmin$ in size with coverage
$\ge90\%$ of the maximum in the central
$\sim 0.5\arcmin \times 1\arcmin$ area.

PACS photometry is measured in 18 apertures with radii from 2 to
50\arcsec\ using the ``annularSkyAperturePhotometry'' task within {\sc
  hipe}, with each measurement corrected for the encircled energy
fractions using PACS responsivity version 7.  The uncertainties
in the flux density measurements are determined from the dispersion in
1000 samples of the total flux in the same number of randomly selected
pixels as included in each aperture, with pixels containing sources or those
with $<50\%$ of maximum coverage excluded from selection.  We then
determine the ``total'' flux density and uncertainty for each target
by fitting a curve of growth to the aperture fluxes and adding 5\%
calibration uncertainty.\footnote{\url{http://herschel.esac.esa.int/twiki/bin/view/Public/PacsCalibrationWeb}} These total flux densities are
presented in Table~\ref{tab:phot}, where we also include PACS
100\,\micron\ data from George et al.\ (in prep.; \herschel\
program OT1\_rivison\_1; see also \citealt{George15})  and the publications
presented in Table~\ref{tab:targs}. \citet{George15} also includes
160\,\micron\ data from OT1\_rivison\_1, although their flux
measurements can be 1--2$\sigma$ lower than those presented
here, since point sources are assumed. 
For HXMM01, the PACS 70 and 160\,\micron\ photometry was independently
reduced and measured in \citet{Fu13}.  Our measurements are consistent
with those results, and we include the \citet{Fu13}
100-\micron\ photometry in the SED fits (Section~\ref{sec:seds}) and
Table~\ref{tab:phot}. H\bootes01 has PACS 100-\micron\ data from
 HerMES GTO time, which are also included here.

\begin{deluxetable}{lcccc}
\tablecaption{PACS 70- and 160-\micron\ photometry and derived far-IR fluxes and luminosities. \label{tab:phot}}
\startdata 
\hline\hline  
Name & $S_{70}$$^{\rm a}$ &  $ S_{160}$  & $L_{\rm FIR}$$^{\rm b}$ & \fir$^{\rm b}$ \\ 
&   (mJy) & (mJy) & ($ 10^{13}~L_{\sun}$) & ($10^{-15}~{\rm Wm^{-2}}$)  \\
\hline 
 G15v2.19 & $ 316\pm 16$  & $1077\pm 55$ &  $  3.2\pm0.3$ & $ 16.6\pm 0.7 $  \\
 G09v1.40 & $<  9$        & $ 279\pm 16$ &  $  4.6\pm0.3$ & $  4.0\pm 0.1 $  \\
 G12v2.257 & $  15\pm  4$ & $ 147\pm 11$ &  $  1.7\pm0.3$ & $  1.3\pm 0.1 $  \\
 GP.NA.144 & $  11\pm  3$ & $ 177\pm 12$ &  $  4.0^{+ 0.4}_{-0.3}$ & $  3.1\pm 0.1 $  \\
 NGP.NA.56 & $  14\pm  3$ & $ 303\pm 18$ &  $  7.2^{+ 0.4}_{-0.3}$ & $  5.0\pm 0.1 $  \\
   HXMM01 & $  10\pm  3$  & $ 123\pm 12$ &  $  2.9\pm0.3$ & $  2.0\pm 0.1 $  \\
 G09v1.124 & $  16\pm 4$  & $ 169\pm 11$ &  $  4.4\pm0.3$ & $  2.7\pm 0.1 $  \\
 G15v2.235 & $< 11$       & $ 115\pm  9$ &  $  3.6\pm0.3$ & $  2.1\pm 0.1 $  \\
 G09v1.326 & $<  8$       & $ 126\pm  9$ &  $  3.1\pm0.4$ & $  1.6\pm 0.1 $  \\
 NGP.NB.78 & $  40\pm  3$ & $ 210\pm 12$ &  $  8.2\pm0.7$ & $  2.7\pm 0.1 $  \\
 G12v2.43 & $  16\pm  3$  & $ 219\pm 12$ &  $  8.9^{+ 0.7}_{-0.6}$ & $  2.8\pm 0.1 $  \\
 G12v2.30 & $  30\pm  4$  & $ 228\pm 13$ &  $ 11.2^{+ 0.7}_{-0.6}$ & $  3.3\pm 0.1 $  \\
 HBo\"otes01 & $<  4$     & $  81\pm  6$ &  $  5.6\pm0.5$ & $  1.7\pm 0.1 $ 
\enddata 
\tablecomments{
All measurements are apparent values (i.e.\ no corrections have been
made for the lensing amplification).
$^{\rm a}$ $3\sigma$ upper limits are presented 
for undetected sources.
$^{\rm b}$ $L_{\rm FIR}$ and \fir, measured from the modified
blackbody fits in Figure~\ref{fig:seds} are the far-IR
luminosity (40--500\,\micron) and continuum flux
(42.5--122.5\,\micron), respectively (Section~\ref{sec:seds}).
}
\end{deluxetable}

\subsection{Archival sample and data}
\label{sec:archsamp}

To identify additional SMGs with IR spectroscopy we searched the
successful \herschel\
proposals\footnote{\url{www.cosmos.esa.int/web/herschel/observing-overview}}
for those targeting high-redshift star-forming galaxies
(i.e. excluding AGN and QSOs) for PACS spectroscopy. Having identified
likely programs we next searched the \herschel\ Science Archive for
those with SMGs as targets and retained the observations of IR
emission lines that overlap with those studied by our own program
(Section~\ref{sec:spec}). This search resulted in spectroscopy for an
additional 32 SMGs, at CO or optical spectroscopic redshifts of 1.1 to 4.2. Most of these
additional SMGs are gravitationally lensed because the PACS sensitivity
means that only the apparently brightest sources can be observed. These archival
observations covered between one and seven emission lines per galaxy.
The full list of archival targets and data included in our analyses
are presented in Table~\ref{tab:archsamp}.  The archival sources are
broadly consistent with the main SMG population and the individually
targeted galaxies, in terms of the IR-luminosity and redshift distributions,
and with IR emission being dominated by star-formation. This archival sample includes
LESS SMGs \citep{Coppin12}, lensed HerMES and H-ATLAS sources from a
similar followup program to this (OT1\_averma\_1; Verma et al.\ in
prep.), lensed SPT sources \citep{Vieira13}, and other SMGs.

The PACS spectroscopy of the archival targets is reduced in the same
way as the targeted data (Section~\ref{sec:spec}). For those spectra
that have been published elsewhere we have verified that
our reduction produces measurements consistent with the published
data. PACS photometry is not available for most of the archival
targets, so those are not considered here; we instead use the
published IR luminosity of each source, where necessary scaling to the 
wavelength ranges for \lfir\ and \fir\ (Section~\ref{sec:seds}) by using the
SED fits of the targeted sources (Section~\ref{sec:seds}). For sources
with multiple published IR luminosities we use the one
constrained by the most photometric data points.

\begin{figure*}
\centering\includegraphics[width=0.95\textwidth]{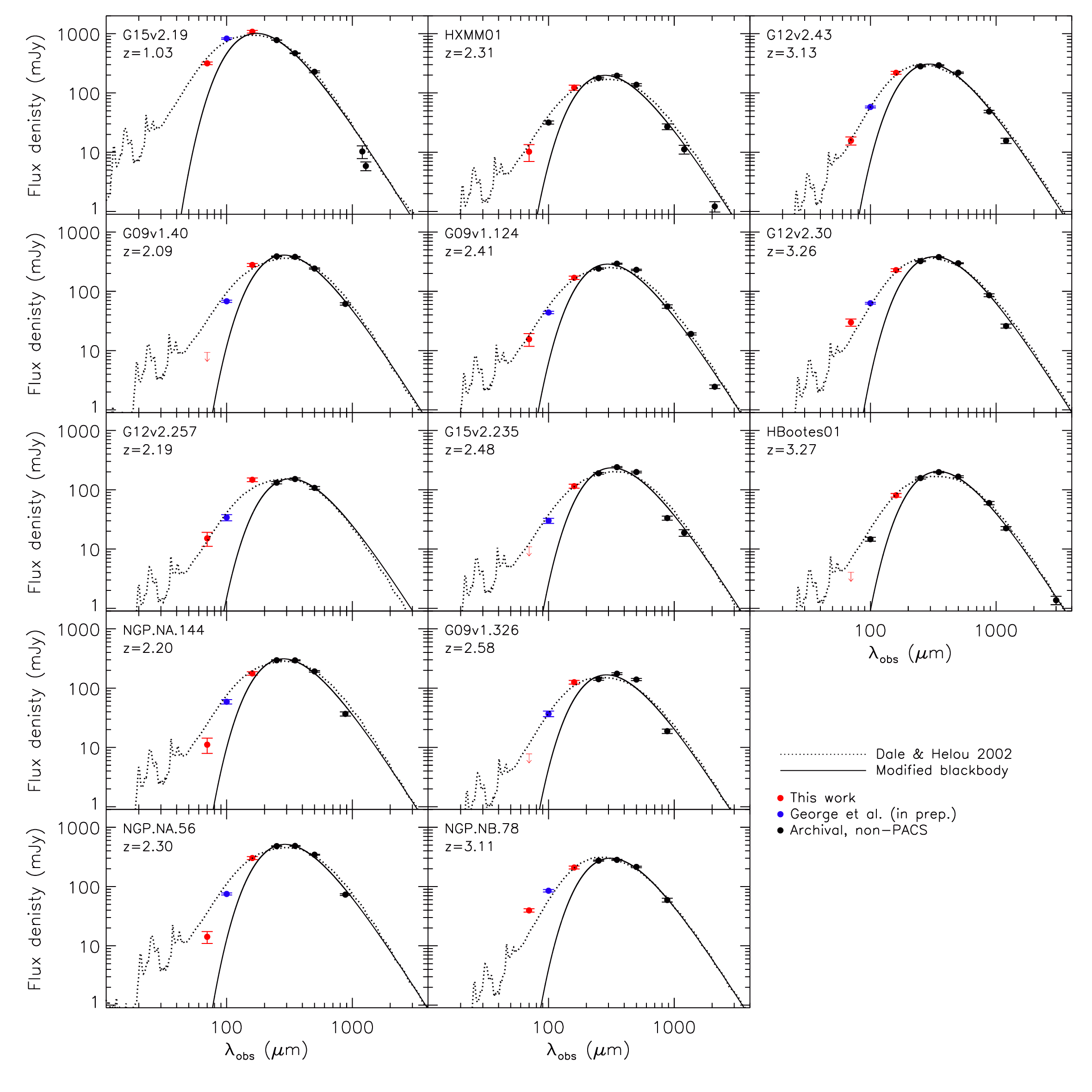}
\caption{Far-IR SEDs of the 13 targets in our sample, combining
  existing photometry (see references in Table~\ref{tab:targs}) with our new
  \herschel-PACS 70 and 160\,\micron\ data and 100\,\micron\
  measurements from George et al.\ (in prep.). The best-fit
  modified black body SEDs,
  which are used to calculate \lfir\ and \fir\ (see
  Section~\ref{sec:seds}) are shown, and for each galaxy we also present the
  best-fit SED from the \citet{DaleHelou02} template library. 
}
\label{fig:seds}
\end{figure*}

\section{Analysis and Discussion} 
\label{sec:analysis}

\subsection{Far-infrared SED fits}
\label{sec:seds}

The PACS photometry, decsribed in Section~\ref{sec:phot}, is
supplemented with the SPIRE \citep{Griffin10} 250-, 350-, and 500-\micron\  data from
HerMES \citep{Roseboom12, Wang14} and H-ATLAS \citep{Valiante16}, and, where available, longer wavelength follow-up
photometry (see references in Table~\ref{tab:targs}). 
We show the far-IR SEDs derived from this
compilation of data in Figure~\ref{fig:seds}. 

For each galaxy we fit the observed far-infared SED with an optically
thin modified blackbody spectrum of the form
\begin{equation}
S_{\nu} \propto \nu^{\beta}B_{\nu}(T_{\rm D}),
\end{equation}
where $S_{\nu}$ is the flux density, $\nu$ is frequency and $\beta$ is
the power law emissivitity index. $B_{\nu}(T_{\rm D})$ is the Planck
function, defined as
\begin{equation}
B_{\nu}(T_{\rm D}) \propto \frac{\nu^3}{e^{h\nu/kT_{\rm D}} -1},
\end{equation}
for a dust temperature, $T_{\rm D}$, and where $h$ and $k$ denote the Planck
and Boltzmann constants, respectively.
We fix $\beta=1.5$, which is consistent with observed values in a 
range of galaxies \citep[e.g.,][]{Hildebrand83, DunneEales01}, and allow
$T_{\rm D}$ and the normalization to vary. The best-fit modified blackbody
curve for each galaxy is shown in Figure~\ref{fig:seds}.

Using these modified blackbody fits we next calculate both
far-IR luminosity ($L_{\rm FIR}$)  and far-IR continuum
flux (\fir) for each SMG. For consistency with existing studies we follow the definitions of \citet{GraciaCarpio11} and \citet{Coppin12} for
these quantities, whereby:
\begin{itemize}
\item $L_{\rm FIR}$ is the luminosity of the rest-frame SED integrated between 40 and 500\,\micron, and; 
\item \fir\ is the luminosity integrated between 42.5 and 122.5\,\micron\ in
the rest-frame, and converted to flux by dividing by $4\pi D_{\rm L}^2$,
where $D_{\rm L}$ is the luminosity distance. 
\end{itemize}
The apparent (i.e., without correction for lensing amplification) values of \lfir\ and \fir\
calculated from the modified blackbody SED fits are listed in
Table~\ref{sec:seds} and are used in the analysis in
the rest of this paper. 

However, since the single temperature modified blackbody can underpredict the
emission on the Wien side of the far-IR dust peak, it is possible
that the $L_{\rm FIR}$ and \fir\ values that we calculate from the modified
blackbody fits are systematically underestimated. To test the magnitude of this effect we also fit
each galaxy with SEDs from the \citet{DaleHelou02} template library;
these fits are also shown in Figure~\ref{fig:seds}.

There is  no significant systematic offset between \lfir\ from the
two fitting methods, with the median ratio of the \citet{DaleHelou02}
to modified blackbody values being 0.99.  There are only three
galaxies with \lfir\ from the \citet{DaleHelou02} SED fits that are
significantly different to the values from the modified blackbody
fits.  These are G15v2.19, G12v2.43 and H\bootes01, which are 15\%
higher and 10\% and 10\% lower for the \citet{DaleHelou02} fits,
respectively. Only
one galaxy has significantly higher \fir\ from the
\citet{DaleHelou02} SEDs than the modified blackbody fits, which is
G12v2.257 with 30\% difference in \fir. However,
five systems -- G15v2.19, G09v1.40, NGP.NA.144, HXMM01 and NGP.NB.78 -- have lower \fir\ for the \citet{DaleHelou02}
SEDs than the modified blackbody fits.  For these six galaxies the
\fir\ for the \citet{DaleHelou02} fits are 70--95\% of the modified
blackbody values. These galaxies would thus be offset upwards by
$\sim25\%$ in Figures~\ref{fig:ratios} and \ref{fig:h2ratios} if we were
to use \fir\ from the \citet{DaleHelou02} SED fits instead of from the
modified blackbody fits.
The typically small differences between \lfir\ and \fir\ from the
\citet{DaleHelou02} and modified blackbody fits are because \lfir\ and
\fir\ are most sensitive to the peak and long wavelength part of the SED,
where the modified blackbody does a good job of fitting the data.
Note that due to the narrow wavelength ranges considered
for \fir\ and \lfir\ and the slightly lower normalization of some of the
\citet{DaleHelou02} fits, having
lower \lfir\ for the \citet{DaleHelou02} fits  compared with the
modified blackbody in some cases is not unexpected.

\begin{figure*}
\centering\includegraphics[width=0.95\textwidth]{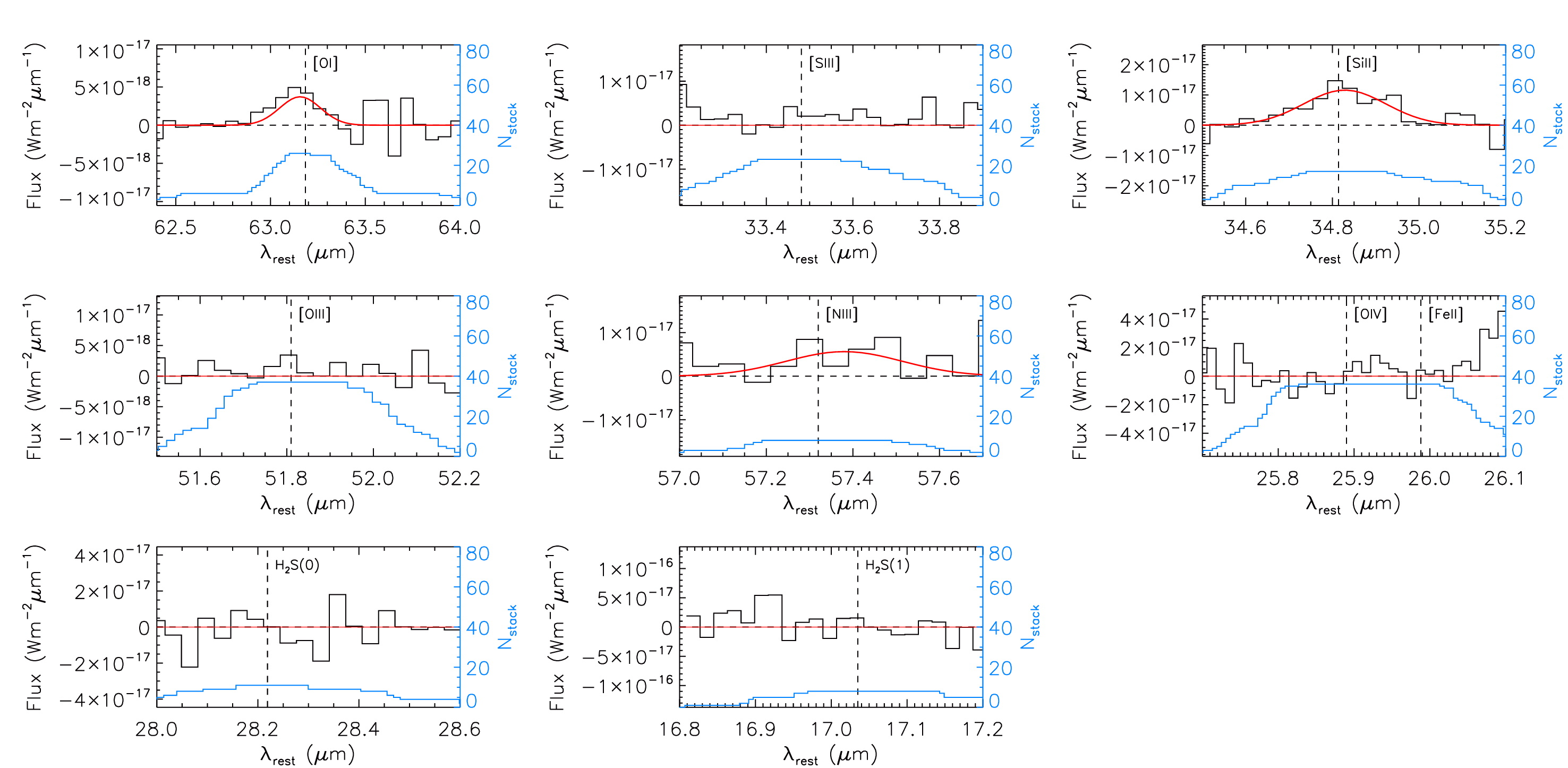}
\medskip\caption{
Mean stacks of the continuum-subtracted rest-frame spectra for each
  of the observed IR emission lines, with the best-fit
  $\ge3\sigma$ Gaussian line profiles
  overlaid (red). The lower (blue) line corresponds to the
  right-hand axis and shows the number of spectra included in each bin of
  the stack, which is variable because of the different rest-frame
  coverage of each observation.
}
\label{fig:linestack}
\end{figure*}

\subsection{Individual emission line measurements}
\label{sec:specmeasure} 

We measure the line fluxes (and upper limits) from the 1D spectra of
the individual galaxies, reduced and extracted as described in
Section~\ref{sec:spec}, and with the noisy regions at the edges of the
spectra (typically 5--10 wavelength bins) removed. Then, with the
exception of the \oiv\ observations, each spectrum is fit with a
single Gaussian line profile and flat continuum component using the
{\sc mpfit} function in {\sc IDL} \citep{Markwardt09}, which uses
non-linear Levenberg-Marquardt least-squares minimization. We constrain
the fits to have non-negative continua and the velocity offsets of the
lines are required to be $\le\pm800~{\rm kms^{-1}}$ from their
expected locations based on the CO redshifts.  The wavelength range of
the \oiv\ observations includes the \feii\ line, and therefore, those
data are fitted with double Gaussians, using the same {\sc mpfit} {\sc
  IDL} function.  In all cases the velocity-integrated flux in each
line is calculated from the continuum-subtracted best-fit Gaussian.

The pipeline-derived uncertainties on the PACS spectra are known to
be unreliable\footnote{PACS Data Reduction Guide for Spectroscopy, \S7.7: \url{http://herschel.esac.esa.int/twiki/pub/Public/PacsCalibrationWeb/PDRGspec_HIPE14p2.pdf}}, and therefore we weigh each wavelength bin equally for fitting purposes.
The uncertainty on the line fluxes are determined from 1000 trials
for each line, wherein we add random noise with the same $1\sigma$ rms
as measured from the line-free portions of the spectra and refit the
line. The $3\sigma$ detection limit for each line is calculated from a
Gaussian profile with a peak height three times the rms noise in the
spectra, centered at the expected position of the emission line from
the CO redshift. For
the purposes of this calculation we assume a linewidth of
300~km\,s$^{-1}$ FWHM, which is consistent with observations of
high-redshift star-forming galaxies \citep[e.g.,][]{Sturm10, Coppin12}
and similar to the PACS instrumental resolution.

The spectra and line fits for the 13 targets of OT2\_jwardlow\_1 are presented in
Appendix~\ref{app:lines} and the measurements
given in Table~\ref{tab:spec}.

\renewcommand{\tabcolsep}{0.2mm}
\begin{deluxetable*}{lccccccccccccc}
\tablecaption{Spectral measurements \label{tab:spec}}
\startdata 
\hline\hline
% \begin{deluxetable*}{lccccccccc} is the setup
   & \multicolumn{3}{c}{\oi\micron} &  \multicolumn{3}{c}{\sxiii\micron} &  \multicolumn{3}{c}{\sixii\micron} \\
Name &  Line flux & Line $\lambda_{\rm obs}$$^a$ & Continuum$^b$  
     &  Line flux & Line $\lambda_{\rm obs}$$^a$ & Continuum$^b$  
     &  Line flux & Line $\lambda_{\rm obs}$$^a$ & Continuum$^b$  \\
   & (10$^{-18}$Wm$^{-2}$) & (\micron) & (mJy) 
   & (10$^{-18}$Wm$^{-2}$) & (\micron) & (mJy) 
   & (10$^{-18}$Wm$^{-2}$) & (\micron) & (mJy) \\
\hline
G15v2.19 & $< 35.3$ & 128.07 & $1044\pm 157$ & \nodata & \nodata & \nodata & \nodata & \nodata & \nodata\\
G09v1.40 & $< 81.0$ & 195.29 & $< 549$ & \nodata & \nodata & \nodata & $< 21.2$ & 107.61 & $<  79$\\
G12v2.257 & \nodata & \nodata & \nodata & $< 10.5$ & 106.84 & $<  39$ & $< 10.9$ & 111.09 & $<  42$\\
NGP.NA.144 & $< 44.4$ & 202.30 & $< 312$ & $< 20.6$ & 107.21 & $<  76$ & $< 19.8$ & 111.47 & $<  76$\\
NGP.NA.56 & $< 34.0$ & 208.62 & $< 245$ & $< 24.5$ & 110.55 & $<  94$ & $< 27.0$ & 114.96 & $< 107$\\
HXMM01 & $< 12.7$ & 208.94 & $<  92$ & $< 13.7$ & 110.72 & $<  52$ & $< 15.6$ & 115.13 & $<  62$\\
G09v1.124 & \nodata & \nodata & \nodata & $< 15.8$ & 114.17 & $<  62$ & $< 17.9$ & 118.72 & $<  73$\\
G15v2.235 & \nodata & \nodata & \nodata & $< 12.8$ & 116.48 & $<  51$ & $< 11.4$ & 121.12 & $<  47$\\
G09v1.326 & \nodata & \nodata & \nodata & $< 13.4$ & 119.96 & $<  55$ & \nodata & \nodata & \nodata\\
NGP.NB.78 & \nodata & \nodata & \nodata & $< 19.1$ & 137.64 & $ 128\pm  91$ & $< 17.1$ & 143.12 & $ 151\pm  84$\\
G12v2.43 & \nodata & \nodata & \nodata & $< 17.3$ & 138.21 & $ 103\pm  82$ & $< 15.3$ & 143.71 & $ 119\pm  76$\\
G12v2.30 & \nodata & \nodata & \nodata & $< 32.4$ & 142.60 & $< 160$ & $< 29.7$ & 148.27 & $< 152$\\
HBootes01 & \nodata & \nodata & \nodata & $< 11.2$ & 143.10 & $<  55$ & $< 12.3$ & 148.80 & $<  63$\\
{\bf Mean Stack$^c$} & $\mathbf{  1.0\pm  0.3}$ & $\mathbf{ 63.16}$ & \ldots & $\mathbf{<  0.9}$ & \nodata & \ldots & $\mathbf{  2.8\pm  0.4}$ & $\mathbf{ 34.83}$ & \ldots \\
\hline\hline
   & \multicolumn{3}{c}{\oiii\micron} &  \multicolumn{3}{c}{\niii\micron} &  \multicolumn{3}{c}{} \\
Name &  Line flux & Line $\lambda_{\rm obs}$$^a$ & Continuum$^b$  
     &  Line flux & Line $\lambda_{\rm obs}$$^a$ & Continuum$^b$  
     &   &  &   \\
   & (10$^{-18}$Wm$^{-2}$) & (\micron) & (mJy) 
   & (10$^{-18}$Wm$^{-2}$) & (\micron) & (mJy) 
   &  &  &  \\
\hline
G15v2.19 & \nodata & \nodata & \nodata & $< 32.9$ & 116.19 & $ 913\pm 132$ &  &  & \\
G09v1.40 & $< 45.4$ & 160.14 & $< 252$ & $< 18.6$ & 177.18 & $ 220\pm 114$ &  &  & \\
G12v2.257 & $< 14.3$ & 165.33 & $<  82$ & \nodata & \nodata & \nodata &  &  & \\
NGP.NA.144 & $ 10.6\pm  3.2$ & 165.86 & $< 134$ & $< 22.8$ & 183.54 & $< 145$ &  &  & \\
NGP.NA.56 & $< 23.2$ & 171.08 & $ 241\pm 137$ & $< 28.5$ & 189.27 & $ 342\pm 187$ &  &  & \\
HXMM01 & $<  7.2$ & 171.34 & $ 129\pm  42$ & \nodata & \nodata & \nodata &  &  & \\
G09v1.124 & $< 27.3$ & 176.67 & $< 167$ & \nodata & \nodata & \nodata &  &  & \\
G15v2.235 & $< 24.5$ & 180.25 & $< 153$ & \nodata & \nodata & \nodata &  &  & \\
G09v1.326 & $< 23.0$ & 185.64 & $< 148$ & \nodata & \nodata & \nodata &  &  & \\
NGP.NB.78 & \nodata & \nodata & \nodata & \nodata & \nodata & \nodata &  &  & \\
G12v2.43 & \nodata & \nodata & \nodata & \nodata & \nodata & \nodata &  &  & \\
G12v2.30 & \nodata & \nodata & \nodata & \nodata & \nodata & \nodata &  &  & \\
HBootes01 & \nodata & \nodata & \nodata & \nodata & \nodata & \nodata &  &  & \\
{\bf Mean Stack$^c$} & $\mathbf{<  0.9}$ & \nodata & \ldots & $\mathbf{  1.9\pm  0.6}$ & $\mathbf{ 57.38}$ & \ldots &  &  &  \\
\hline\hline
   & \multicolumn{5}{c}{\oiv\micron$^d$ and \feii\micron$^d$} &  \multicolumn{3}{c}{} &  \multicolumn{2}{c}{} \\
Name &  [O\,{\sc iv}] flux & [O\,{\sc iv}] $\lambda_{\rm obs}$$^a$ & Continuum$^b$ & [Fe\,{\sc ii}] flux & [Fe\,{\sc ii}] $\lambda_{\rm obs}$$^a$ 
   & &  & 
   & &  &  \\
   & (10$^{-18}$Wm$^{-2}$) &  (\micron) & (mJy) & (10$^{-18}$Wm$^{-2}$) & (\micron)  
   & &  &  
   & &  &  \\
\hline
G15v2.19 & \nodata & \nodata & \nodata & \nodata & \nodata &  &  &  &  &  & \\
G09v1.40 & $< 90.8$ &  80.03 & $< 252$ & $< 91.2$ &  80.33 &  &  &  &  &  & \\
G12v2.257 & $< 51.4$ &  82.61 & $< 147$ & $< 51.6$ &  82.93 &  &  &  &  &  & \\
NGP.NA.144 & $< 66.2$ &  82.90 & $ 226\pm 190$ & $< 66.5$ &  83.21 &  &  &  &  &  & \\
NGP.NA.56 & $<138.8$ &  85.49 & $< 411$ & $<139.3$ &  85.81 &  &  &  &  &  & \\
HXMM01 & $  7.7\pm  2.5$ &  85.70 & $<  66$ & $< 22.6$ &  85.94 &  &  &  &  &  & \\
G09v1.124 & $< 95.9$ &  88.28 & $< 293$ & $< 96.2$ &  88.62 &  &  &  &  &  & \\
G15v2.235 & $< 56.0$ &  90.07 & $< 175$ & $< 56.2$ &  90.41 &  &  &  &  &  & \\
G09v1.326 & $< 62.6$ &  92.76 & $< 201$ & $< 62.8$ &  93.12 &  &  &  &  &  & \\
NGP.NB.78 & \nodata & \nodata & \nodata & \nodata & \nodata &  &  &  &  &  & \\
G12v2.43 & $<  7.0$ & 106.87 & $  53\pm  26$ & $<  7.0$ & 107.28 &  &  &  &  &  & \\
G12v2.30 & $< 10.7$ & 110.27 & $  56\pm  40$ & $< 10.7$ & 110.68 &  &  &  &  &  & \\
HBootes01 & \nodata & \nodata & \nodata & \nodata & \nodata &  &  &  &  &  & \\
{\bf Mean Stack$^c$} & $\mathbf{<  4.0}$ & \nodata & \nodata & $\mathbf{<  4.0}$ & \nodata & &  &  &  &  &  & \\
\hline\hline
   & \multicolumn{3}{c}{\htsz} &  \multicolumn{3}{c}{\htso} & \multicolumn{3}{c}{}  \\
Name & Line flux & Line $\lambda_{\rm obs}$$^a$ & Continuum$^b$ 
     & Line flux & Line $\lambda_{\rm obs}$$^a$ & Continuum$^b$ 
     &    &  &  \\
   & (10$^{-18}$Wm$^{-2}$) & (\micron) & (mJy) 
   & (10$^{-18}$Wm$^{-2}$) & (\micron) & (mJy) 
   & &  &  \\
\hline
G15v2.19 & $< 94.0$ &  57.20 & $< 186$ & \nodata & \nodata & \nodata &  &  & \\
G09v1.40 & $< 31.7$ &  87.22 & $<  96$ & \nodata & \nodata & \nodata &  &  & \\
G12v2.257 & \nodata & \nodata & \nodata & \nodata & \nodata & \nodata &  &  & \\
NGP.NA.144 & \nodata & \nodata & \nodata & $< 69.9$ &  54.55 & $< 132$ &  &  & \\
NGP.NA.56 & $< 41.5$ &  93.18 & $< 134$ & $< 83.2$ &  56.25 & $< 162$ &  &  & \\
HXMM01 & \nodata & \nodata & \nodata & $< 44.4$ &  56.33 & $<  86$ &  &  & \\
G09v1.124 & \nodata & \nodata & \nodata & $< 54.2$ &  58.09 & $< 109$ &  &  & \\
G15v2.235 & \nodata & \nodata & \nodata & $< 31.1$ &  59.26 & $<  64$ &  &  & \\
G09v1.326 & \nodata & \nodata & \nodata & \nodata & \nodata & \nodata &  &  & \\
NGP.NB.78 & $< 10.8$ & 116.01 & $ 107\pm  43$ & \nodata & \nodata & \nodata &  &  & \\
G12v2.43 & $<  9.9$ & 116.49 & $  86\pm  40$ & \nodata & \nodata & \nodata &  &  & \\
G12v2.30 & $< 16.0$ & 120.18 & $<  66$ & $< 34.2$ &  72.55 & $<  86$ &  &  & \\
HBootes01 & $<  6.2$ & 120.61 & $<  26$ & \nodata & \nodata & \nodata &  &  & \\
{\bf Mean Stack$^c$} & $\mathbf{<  3.8}$ & \nodata & \nodata & $\mathbf{<  4.2}$ & \nodata & \nodata &  &  &  

\enddata
\tablecomments{
$3\sigma$ upper limits are given for lines that are not detected above
the $3\sigma$ significance level. 
Parameters for lines without observations are left blank. 
$^a$ For detected lines the wavelength corresponds to
the measured (observed frame) position of the line, otherwise 
 the expected (observed frame) wavelength is given, based on the
 nominal redshifts in Table~\ref{tab:targs}.
$^b$ The continuum flux measured adjacent to the emission line.
$^c$ The stack values are discussed in Section~\ref{sec:linestack}
$^d$ The \oiv\ and \feii\ lines occur close together in a single
spectrum and are therefore fit simultaneously.
}
\end{deluxetable*}
\renewcommand{\tabcolsep}{3mm}

\subsection{Stacked spectra}
\label{sec:linestack}

We next investigate the average properties of the spectra by stacking
the observations of each transition for all the galaxies, which
reduces the background noise by a factor of $\sim\sqrt{N}$ for a stack
of $N$ galaxies with the same background. To trace to
fainter noise limits we include both the 13 targeted galaxies, and the
32 archival sources in the stacks. We have verified that the measured
line fluxes (or limits) are consistent whether or not the archival
data are included. For each line the stacked spectra  contains
 8--37 galaxies and we therefore expect improvements of
factors of $\sim3$--6 in the average sensitivity  of individual
spectra by stacking.

To perform the stacking we first shift each spectrum to the rest frame
and subtract the continuum. We generate a base rest-frame wavelength
grid with spacing equal to the average rest-frame native PACS
resolution for each line targeted. The individual spectra are then rebinned
to the new rest-frame wavelength grid and three different stacks are
generated:
\begin{itemize}
\item Our fiducial method is a mean stack, derived by calculating the
  mean value in each wavelength bin. These mean spectra for each
  targeted emission line are shown in Figure~\ref{fig:linestack} and
  the measurements are presented in Table~\ref{tab:spec}. For
  the line flux/\fir\ ratios examined in Sections~\ref{sec:ratios} and
  \ref{sec:model} we use the mean \fir\ (42.5--122.5\,\micron) of the sources
  included in the stack, such that the ratio is equivalent to
  mean(line flux)/mean(\fir). 
\item We also generate median stacks, consisting of the median value in each
  wavelength bin, which are used to investigate whether a few bright
  outliers dominate the fiducial mean stacks. 
\item To investigate the presence of trends with infrared emission
  weighted mean stacks are also produced, where each source is
  weighted by $1/{\it FIR}$ (42.5--122.5\,\micron), In this case measurements
  from the weighted mean stacks are equivalent to mean(line
  flux/\fir).
\end{itemize}

The rest-frame wavelength coverage from different observing programs
varies, so the number of galaxies contributing to each wavelength bin
varies, as is shown in Figure~\ref{fig:linestack}.  In each case the
stacked spectra are fit using the methodology described in
Section~\ref{sec:specmeasure} for the individual observations. Since
the number of data points stacked in each wavelength bin varies, the
noise level is weighted across the spectra according to
$1/\sqrt{N}$. This is valid since the observations for each line have
similar depths. For the fiducial mean stacks the measured average
fluxes (or limits) are reported in Table~\ref{tab:spec} and the
line/\fir\ (42.5--122.5\,\micron) ratios in
Table~\ref{tab:stackratio}.

The measured linewidths for the \oi, \sixii, and \niii\ (those
with $\ge3\sigma$ detections) in the fiducial mean
stacks  are $510\pm160$, $820\pm280$,
and $700\pm240\,{\rm kms^{-1}}$, respectively
(Figure~\ref{fig:linestack}).  We caution that these apparent
linewidths are not physically meaningful, since they include a
contribution from for potential offsets between the literature
spectroscopic redshifts (due to broadband CO searches or optical data;
Sections~\ref{sec:samp} and \ref{sec:archsamp}) and the targeted IR transitions, which will
artificially broaden the lines.

The line fluxes and upper limits are consistent between the median and
(fiducial) mean stacks, which demonstrates that the mean stacks are
not dominated by a few bright outliers. With the exception of the \oi\
line, the weighted stack measurements are also consistent with the
mean stacks, showing that for most of the lines there is no evidence
of correlations with infrared emission for SMGs. For \oi\ there is
no detection in the weighted stack, with a $3\sigma$ detection limit
of \ois/\fir$<3\times10^{-3}$ (compared with
\ois/\fir$=(3.6\pm1.2)\times10^{-4}$ for the fiducial mean
stack). This suggests that for SMGs there may be a inverse correlation
between infrared luminosity and \oi\ emission. The rest of this paper
focuses on the fiducial mean stacked fluxes, but where relevant we
discuss how the conclusions would change if we instead considered the
\oi\ weighted stack measurement.

\subsection{Individual line strengths}
\label{sec:ratios}

\begin{figure*}
\centering
\begin{tabular}{cc}
\vspace{-0.5cm}
\includegraphics[width=8.5cm]{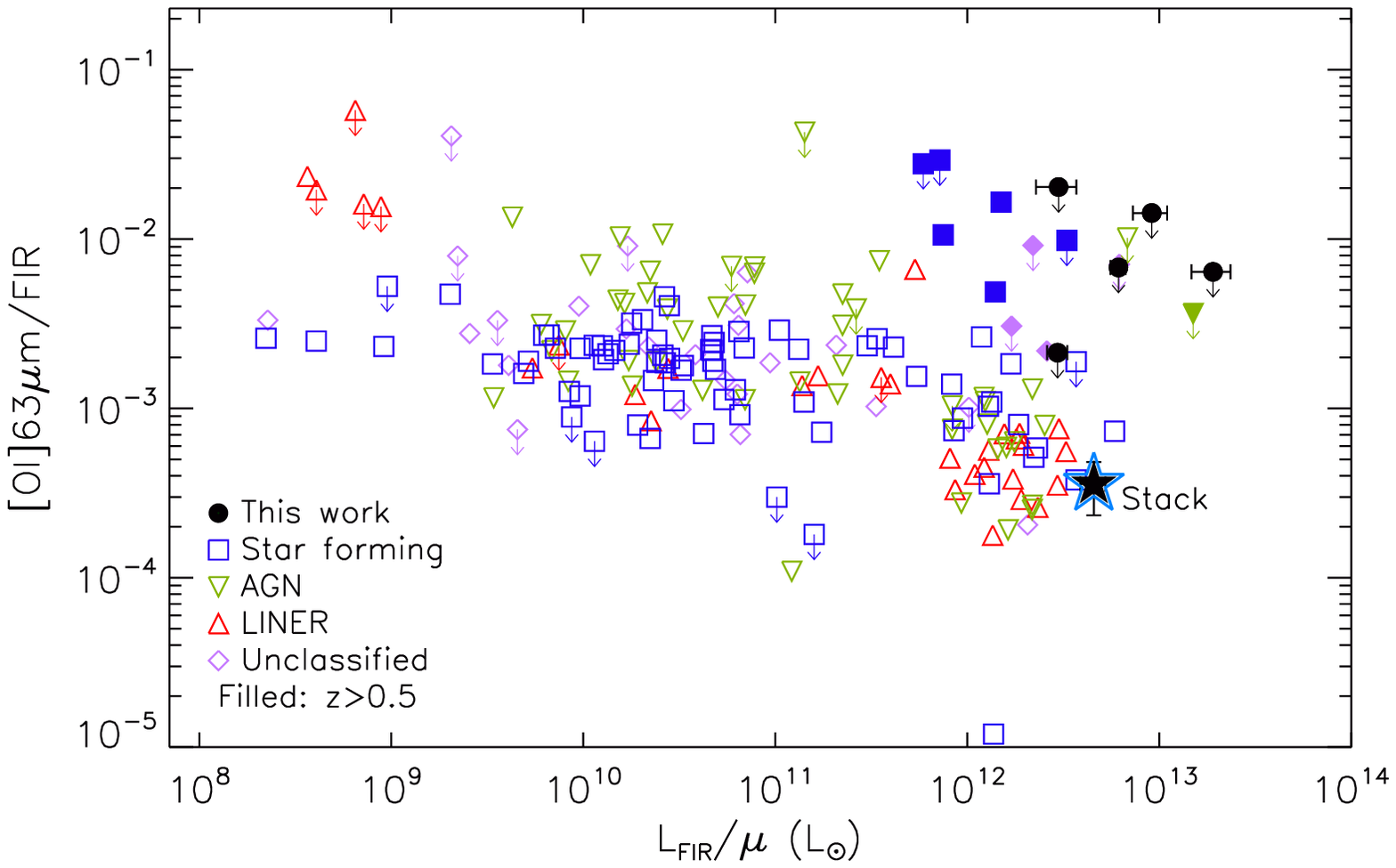} & 
\includegraphics[width=8.5cm]{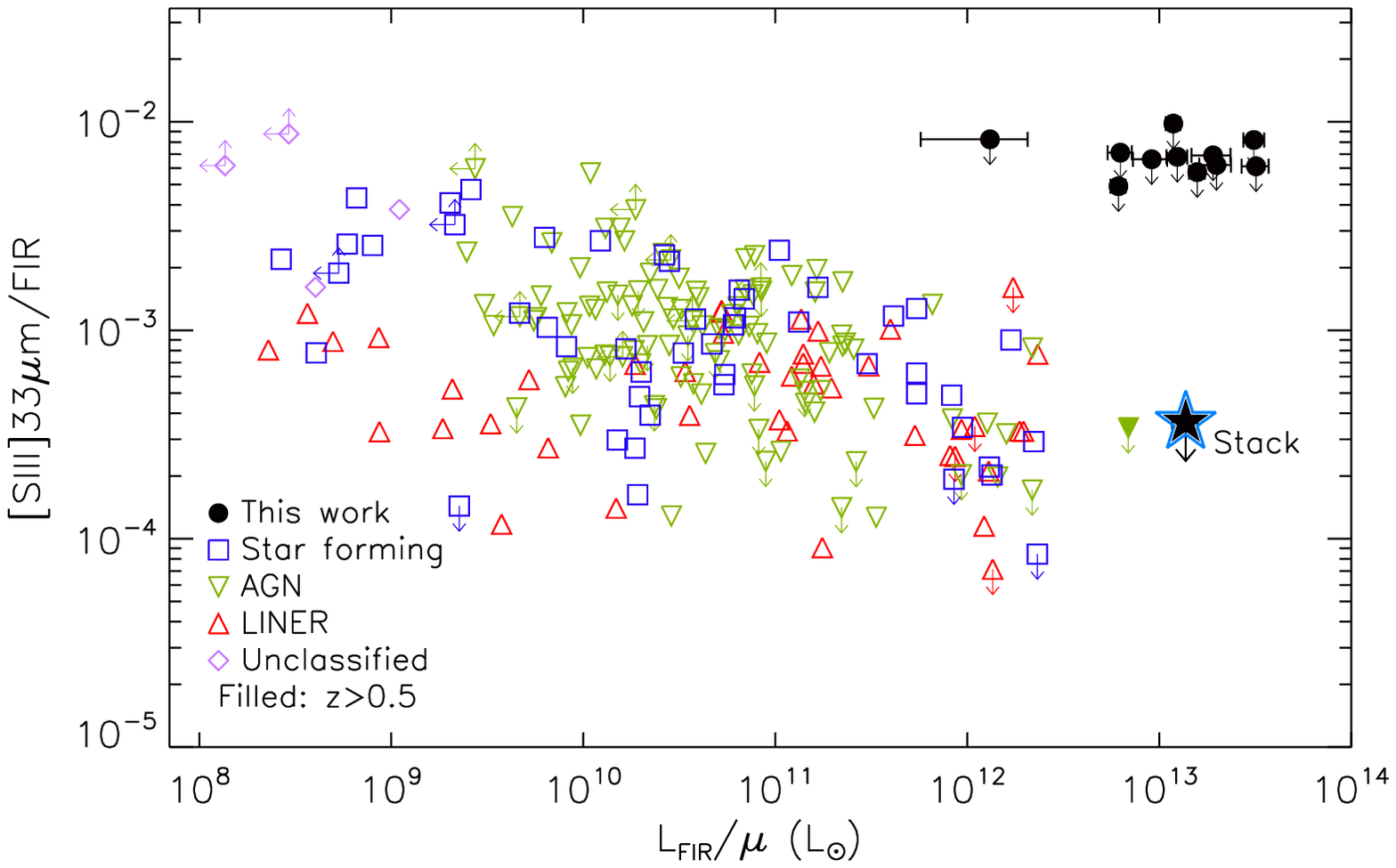} \\\vspace{-0.5cm}
\includegraphics[width=8.5cm]{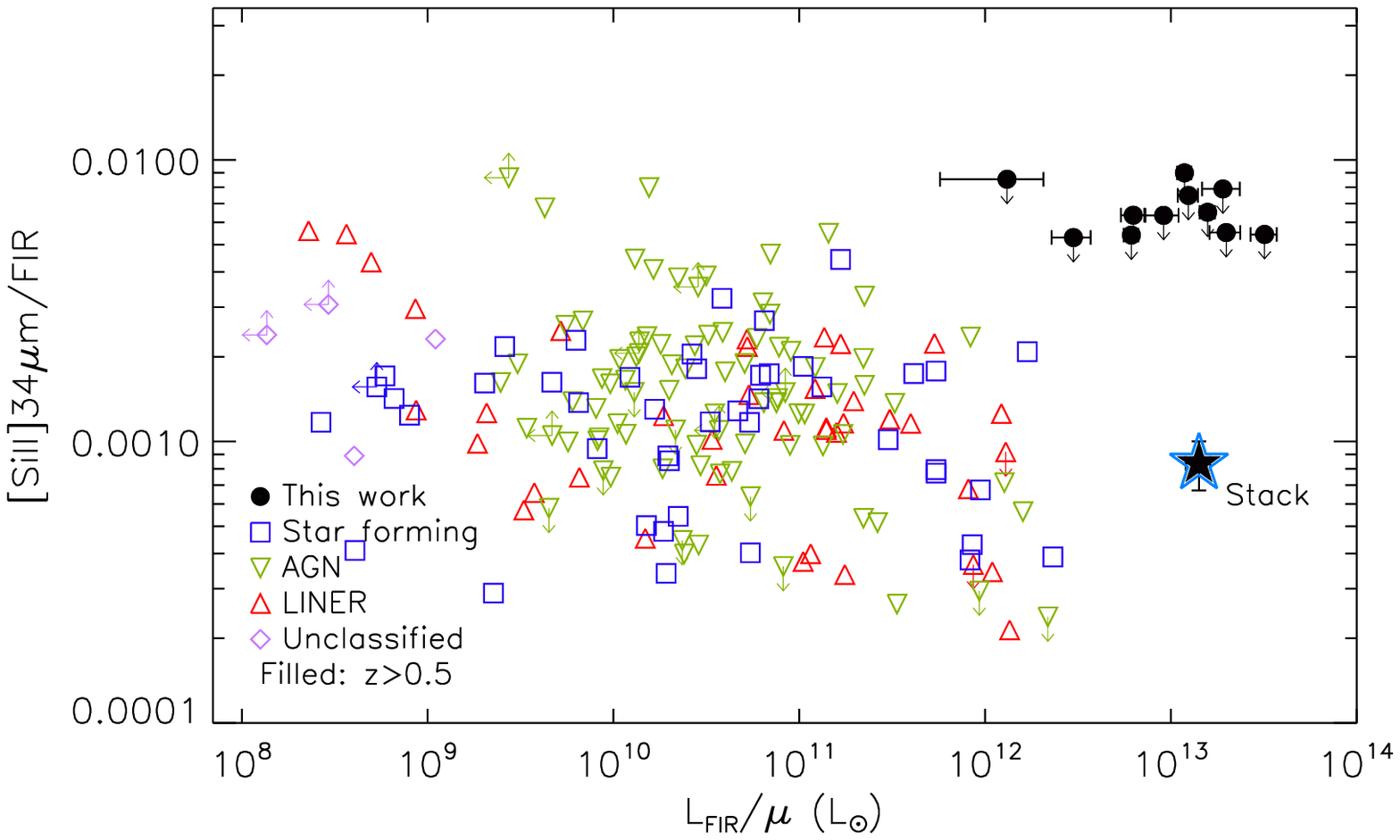} & 
\includegraphics[width=8.5cm]{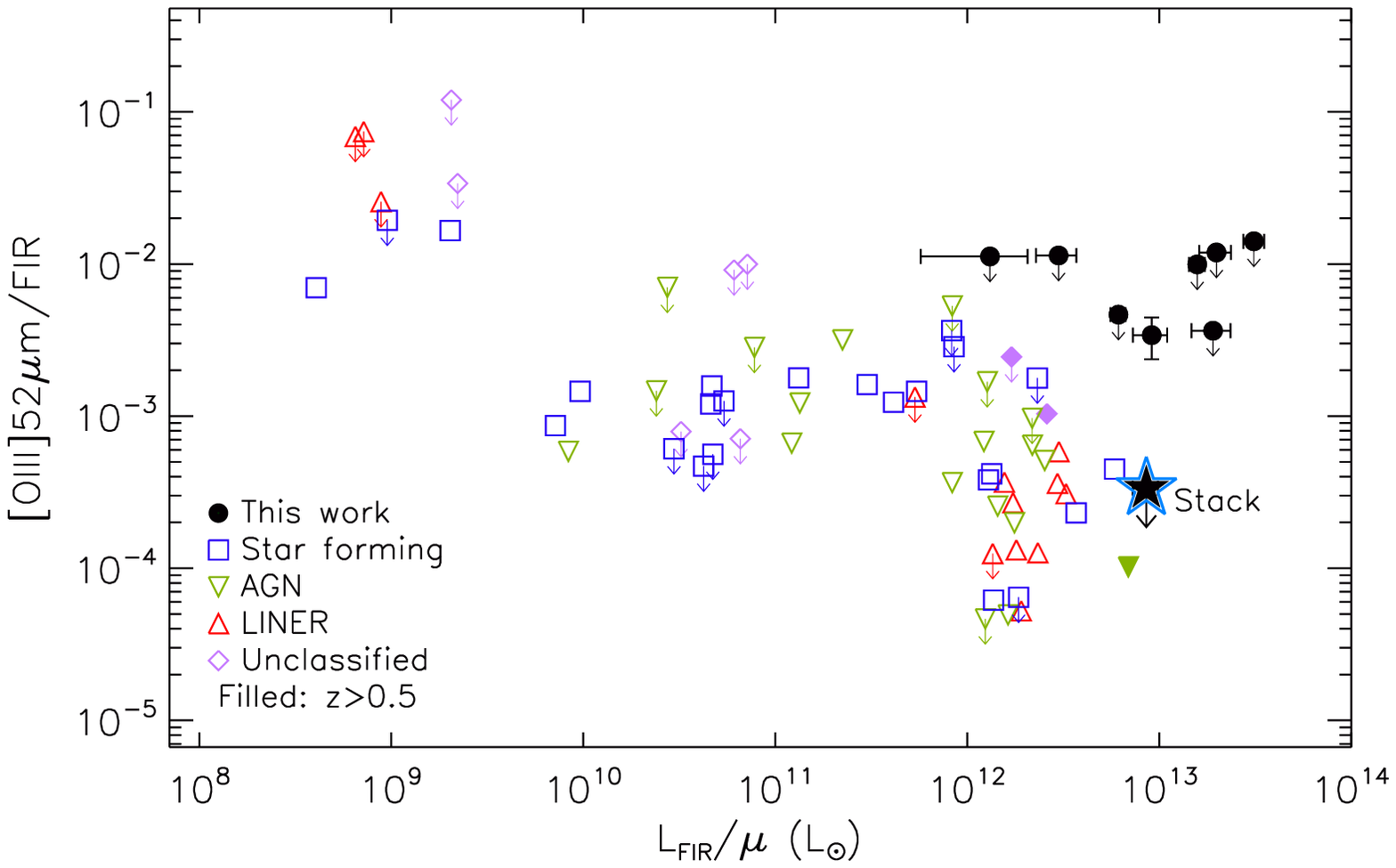} \\\vspace{-0.5cm}
\includegraphics[width=8.5cm]{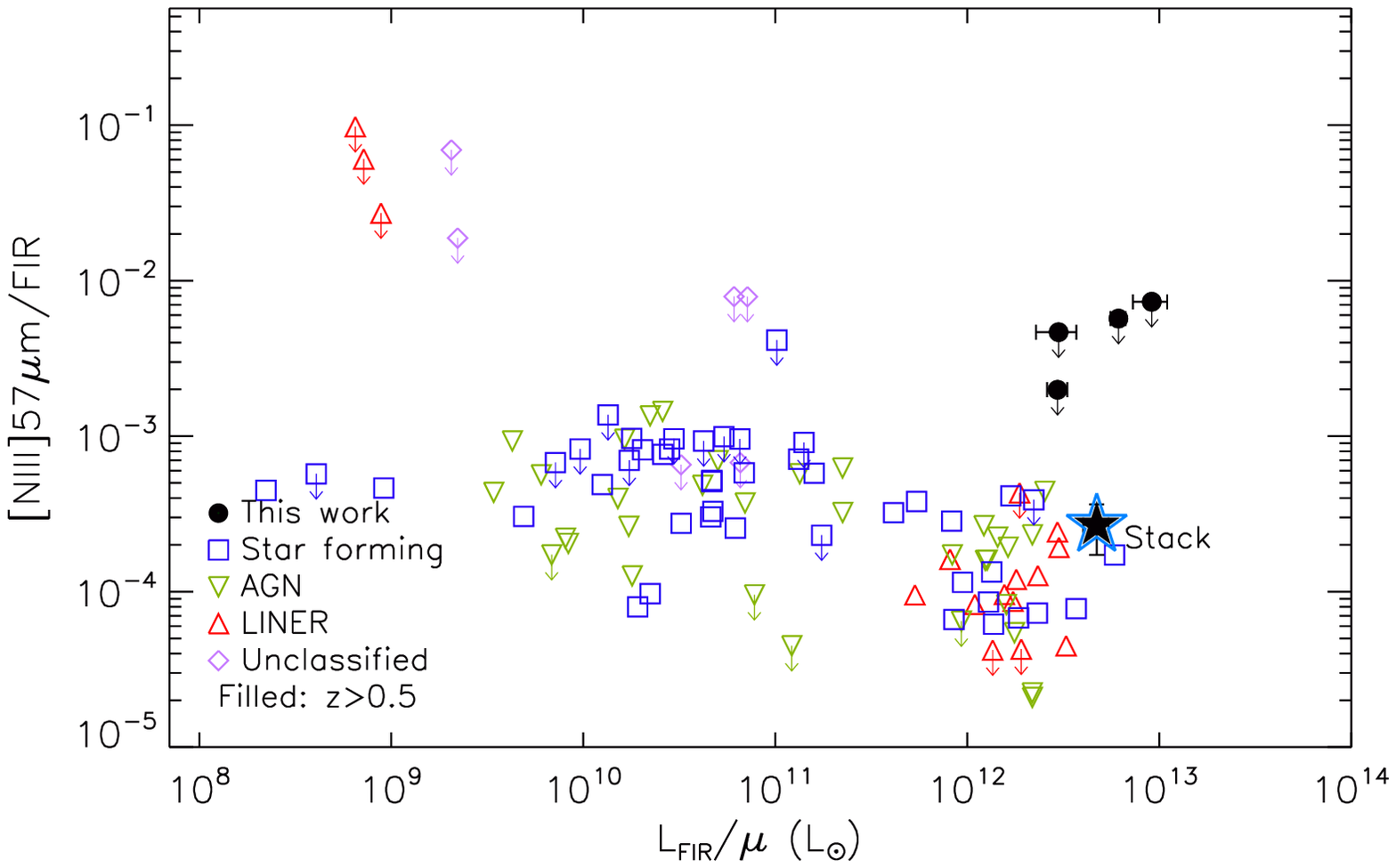} & 
\includegraphics[width=8.5cm]{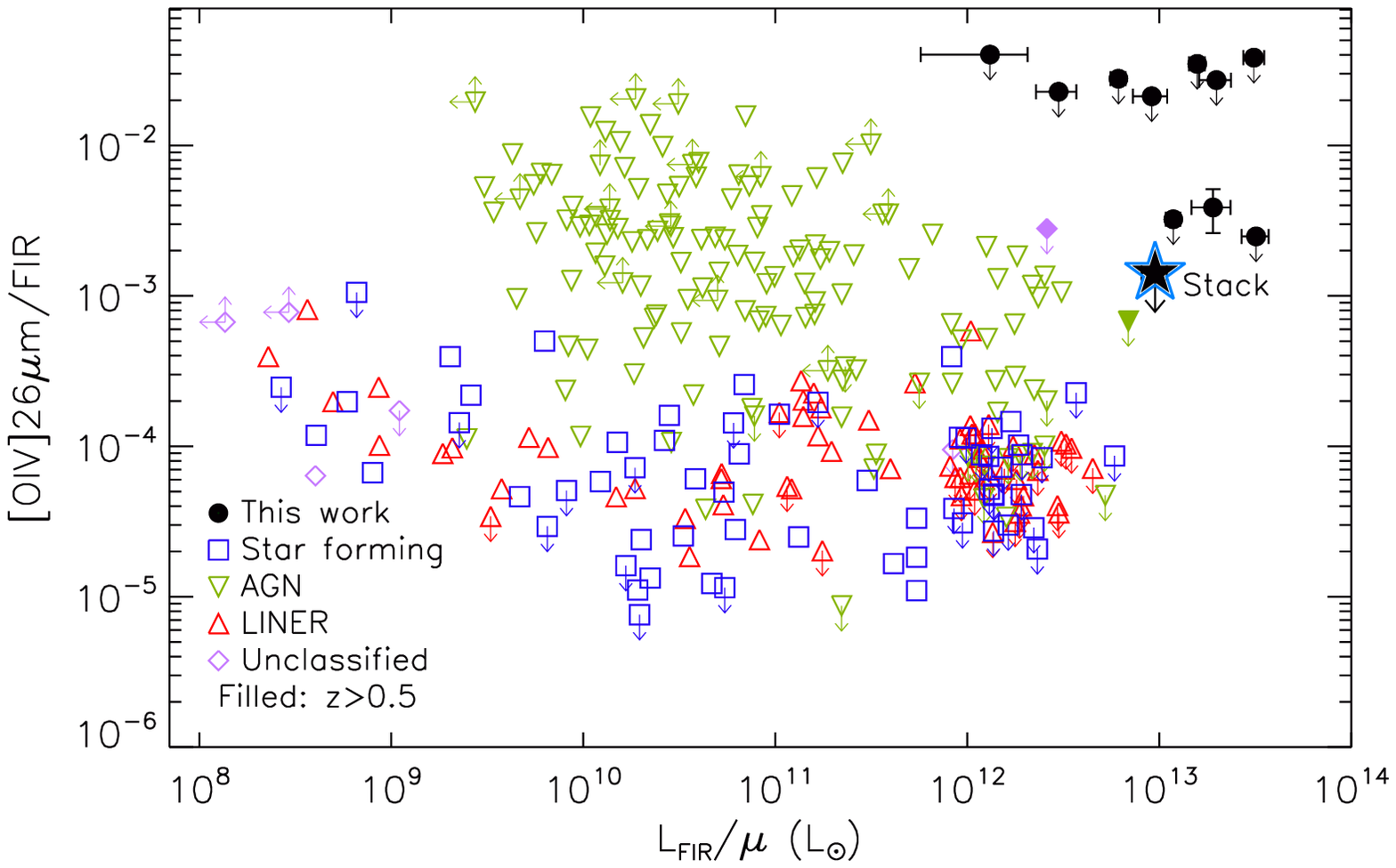}\\\vspace{-0.3cm}
\end{tabular}
\includegraphics[width=8.6cm]{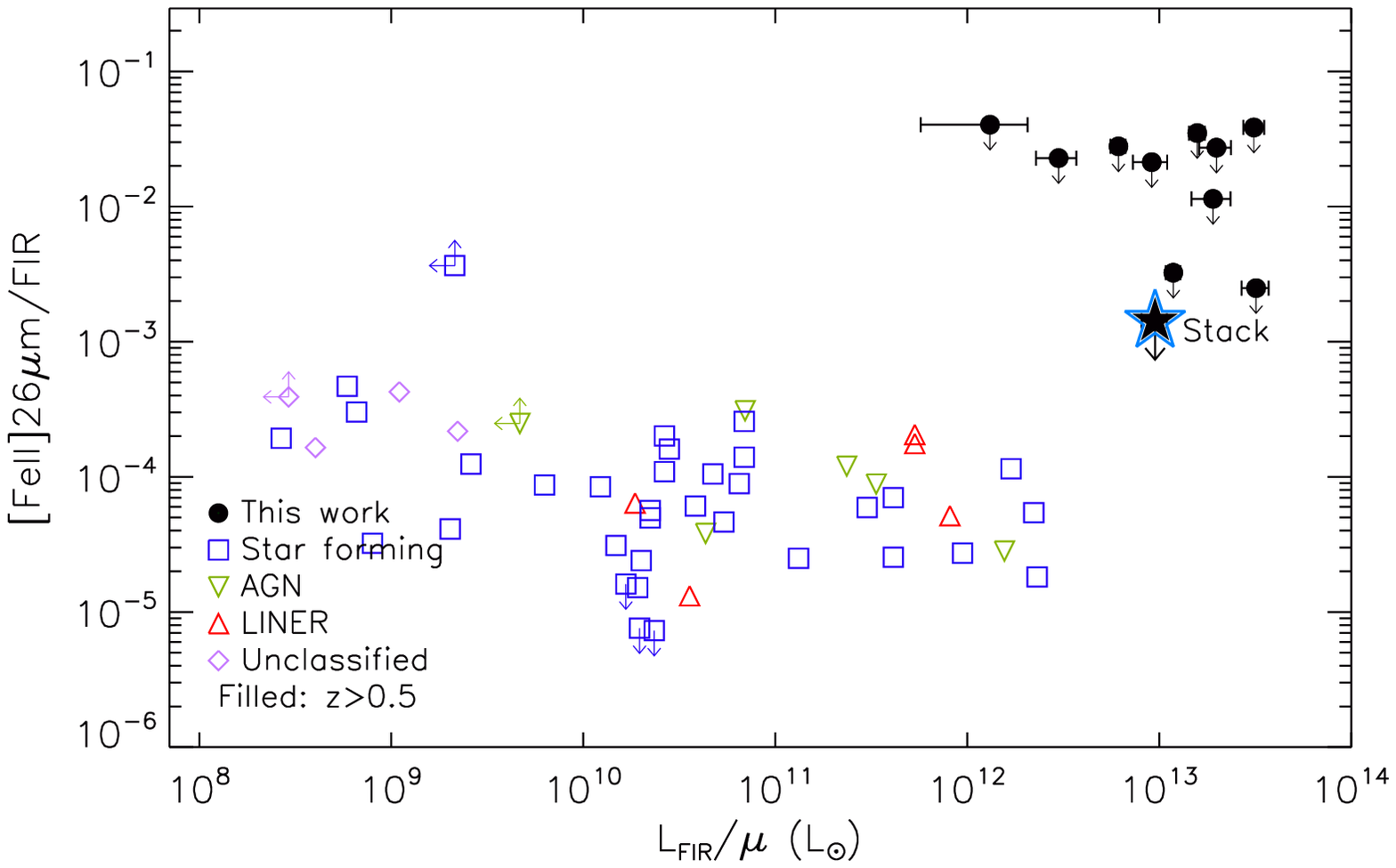} 
\caption{Emission line to continuum \fir\ flux (42.5--122.5\,\micron) as a function of
  (lensing corrected) \lfir\ (40--500\,\micron) for the observed fine structure
  lines. Our individual SMG targets are shown in black,
  with $3\sigma$ upper limits plotted for transitions that are not
  measured above this level. For individual observations
error bars represent $1\sigma$
  uncertainties,  including contributions from the line fitting, from the
  \lfir\ and \fir\ measurements, and from the lensing
amplification. The star represents results from mean stacking
  (Section~\ref{sec:linestack}), using the average \lfir\ and \fir\ of the galaxies
  included in each stack. 
For comparison colored symbols show a compilation of
  star-forming galaxies, AGN, LINERs and unclassified galaxies (see
  Section~\ref{sec:ratios}), with the 
  filled symbols representing those at $z>0.5$ \citep{Farrah07,
    OHalloran08, BernardSalas09, Sturm10, GraciaCarpio11, Coppin12}.
 }
\label{fig:ratios}
\end{figure*}

One way to characterize the strength of IR emission lines is via the
line to \fir\ (42.5--122.5\,\micron) ratio \citep[e.g.,][]{Fischer10,
  Sturm10, GraciaCarpio11, Coppin12, Magdis14}. These are presented in
Figures~\ref{fig:ratios} and \ref{fig:h2ratios} for the fine structure
lines and \htwo\ lines, respectively, and discussed here.
Measurements of the mean stacked spectroscopy for our lensed SMGs are
presented, with the relevant \lfir\ and \fir\  calculated as the mean of
the galaxies included in the stack. These derived average line flux to
\fir\ ratios for SMGs are given in Table~\ref{tab:stackratio}.
Published measurements for other galaxies (mostly at low redshift) are
also shown in Figures~\ref{fig:ratios} and \ref{fig:h2ratios}, colour
coded by whether they are star-forming galaxies, AGN, LINERs or
unclassified \citep{Colbert99, Malhotra01, Negishi01, Sturm02,
  Sturm06, Sturm10, Lutz03, Verma03, Dale04, Farrah07, Farrah13,
  Brauher08, OHalloran08, Tommasin08, Tommasin10, BernardSalas09,
  Hao09, Veilleux09, Hunt10, Ivison10, GraciaCarpio11, Valtchanov11,
  Coppin12, Stierwalt14}. Note that most of the
targeted emission lines (\oi, \sxiii, \oiii\ and \niii) predominantly trace PDRs and \hii\ regions,
and any weak AGN contribution will decrease the
relative line to \fir\ ratio, as the continuum emission is
preferentially enhanced. Whilst energetically dominent or very powerful
AGN can sometimes contribute to the line flux, such
AGN are exceptionally rare in SMGs (e.g. Sections~\ref{sec:oiv} and
\ref{sec:agnsis}; \citealt{Alexander05, Valiante07, Pope08,
  MenendezDelmestre09, Laird10, Wang13}) and  are unlikely to affect
our measurements.

Locally, the relative strength of many PDR cooling lines, including
\oi, \sxiii, \nii\ and \cii, are suppressed with respect to the far-IR
emission in the most luminous systems, particularly those with
`warmer' infrared colours \citep[e.g.,][]{Malhotra01, GraciaCarpio11,
  Farrah13}. Various explanations for the emission line deficits in
high luminosity galaxies have been proposed, including their being
dustier and having higher ionization parameters in the ISM, resulting
in a higher fraction of the UV photons being absorbed by dust and
re-emitted in the far-IR, enhancing the far-IR brightness, and thus
decreasing the line/\fir\ ratios \citep[e.g.,][]{Luhman03,
  GonzalezAlfonso08, Abel09, GraciaCarpio11, Farrah13, Fischer14}.
Alternative explanations include non-PDR flux in the far-IR, such as
from AGN, which would also serve to dilute the PDR line emission
\citep[e.g.,][]{Malhotra01, Luhman03, Farrah13}, or the primary gas
coolant not being the typical \cii\ or \oi\ lines but instead via
other mechanisms \citep[e.g.,][]{Farrah13}. 
It is therefore probable that PDR line deficits may be indicative of a
different `mode' of star formation in local ULIRGs compared with
sub-LIRGs, with the ULIRGs' star-formation being more concentrated, as
is typical in merger-induced activity. We next probe whether SMGs
exhibit similar deficits on a transition-by-transition basis.

\subsubsection{\oi\micron}
\label{sec:oi}

\oi\ has a critical density of $\sim5\times10^5\,{\rm cm}^{-3}$, traces
dense molecular gas with ${\rm T}>100$\,K, and is one of the dominant
cooling lines in dense PDRs. As can be seen in Figure~\ref{fig:ratios},
local LIRGs and ULIRGs have an \oi\ deficit compared with lower
luminosity systems, which typically have \ois/\fir$_{(42.5-122.5\,\micron)}\sim2\times10^{-3}$
\citep[e.g.,][]{Luhman03, GraciaCarpio11, Farrah13}.

Initial observations and publications hinted that \ois/\fir\ may be
enhanced in SMGs, with \ois/\fir\ similar to local sub-LIRGs
\citep{Sturm10, Coppin12}, implying that they may have both large
reservoirs of dense gas (to fuel the far-IR luminosities, and probed
with CO) like local ULIRGs, but with star-formation efficiencies
comparable to late-type galaxies. However, the additional data from
our observations (Figure~\ref{fig:ratios}) now show that the picture
is more complicated, with only four of all 15 SMGs ever observed in \oi\
(MIPS\,J142824.0$+$352619; \citealt{Sturm10}, and unlensed examples
from \citealt{Coppin12}) detected (although mostly at low
significance). The flux limits available for the majority of the
remainder of individual galaxies are not deep enough to provide robust
constraints, leaving interpretation of those results open to
discussion.

Our mean stacked data are significantly more constraining, providing a
$3.2\sigma$ detection, with
\ois/\fir$=(0.36\pm0.12)\times10^{-3}$. Thus, the mean stack result
indicates that, on average, high redshift SMGs behave like local ULIRGs,
having a deficit in their \oi\ emission relative to \fir, although
there are exceptions. The non-detection of \oi\ in the \fir-weighted
stack also suggests that there may be a trend in the strength of \oi\
emission with \fir\ for SMGs.  The difference between the \ois/\fir\
in the few individually detected SMGs (from \citealt{Sturm10} and
\citealt{Coppin12}) and the mean stack is significant, suggesting
there may be physical differences between them, perhaps including a
range of possible \oi\ and \fir\ emission mechanisms for the SMGs. The
two galaxies with the most compelling detections in \citet{Coppin12}
both potentially contain AGN (see their discussion), which may be an
explanation, as weak AGN can strengthen the relative line flux (notice
that the local sources with the highest \ois/\fir\ are AGN and
LINERs).  We further investigate the \oi\ emission in
Section~\ref{sec:model}, where we use PDR models to probe the state of
the ISM in high-redshift SMGs.

\subsubsection{\sxiii\micron}

\sxiii\ has a lower critical density than \oi\ and is a key coolant of
\hii\ regions. Figure~\ref{fig:ratios} shows that locally the
\sxiiis/\fir$_{(42.5-122.5\,\micron)}$ ratio is anti-correlated with
\lfir\ (40--500\,\micron) for star-forming galaxies and AGN. This may
be an effect of continuum dilution of the line, as is observed in
other fine-structure lines \citep[e.g. \cii; ][]{Dale06}.  None of our
individual SMGs or the stack have $\ge3\sigma$ detections in \sxiii.
The upper limits for the undetected individual SMGs are consistent
with luminous local galaxies. The limit on the average \sxiiis/\fir\
of SMGs from the stacked data is lower than many local systems, but
consistent with expectations if the local trend is extrapolated to the
higher luminosity of the stack. Our stacked data are also consistent
with the one other observation of \sxiii\ in a high redshift source
(IRASF10214+4724; \citealt{Sturm10}).

\subsubsection{\sixii\micron}
\label{sec:sixii}

\sixii\ is an important cooling line in PDRs. It has a higher
ionization energy than \oi, but a similar critical density; thus
higher intensity radiation fields are required to excite \sixii\
compared to \oi, and it can also be emitted from XDRs.  Observations
of \sixii\ have so far been limited to the local Universe, where
\sixiis/\fir$_{(42.5-122.5\,\micron)}\sim10^{-3}$ for AGN, star-forming galaxies and
LINERs. Locally, there is hint of a trend of lower \sixiis/\fir\ in
the highest luminosity systems, but not as convincingly as for \oi\ or
\cii\ (for example). There are also indications that the highest
\sixiis/\fir\ values ($\gtrsim4\times10^{-3}$) are only present in
LINERs and AGN, i.e.\ high ionization environments
(Section~\ref{sec:agnsis}).  None of the 11 individual SMGs targeted here
are detected at the $\ge3\sigma$ level. However, the \sixii\ stack is
our strongest detection of all the data (7.5$\sigma$), and shows that
on average SMGs have \sixiis/\fir$=(8.4\pm1.7)\times10^{-4}$. This
average value of \sixiis/\fir\ for SMGs is marginally lower than the
local average for all galaxies, but substantially higher than may be
expected if the weak trend of \sixii/\fir\ dropping for high
luminosity local galaxies continues into the ULIRG regime.

\subsubsection{\oiii\micron}

\oiii\ is an efficient tracer of \hii\ regions and local ULIRGs show
the same suppression in \oiiis/\fir$_{(42.5-122.5\,\micron)}$ as for many of the other
fine-structure lines.  In our \oiii\ observations one SMG is detected
at $>3\sigma$ (NGP.NA.144); the remaining seven individually targeted
SMGs and the stack are undetected. In all cases our constraints on
\oiiis/\fir\ are consistent with the local LIRG and sub-LIRG
population, although in most cases we cannot rule out lower ratios for
SMGs (e.g.\ as is observed in F10214 at $z=2.28$;
\citealt{Sturm10}). The detected SMG, NGP.NA.144, has
\oiiis/\fir~$=(3.4\pm1.0)\times10^{-3}$, which is significantly higher
then local ULIRGs, but is somewhat consistent with local LIRGs and
sub-LIRGs. In contrast, the stack has
\oiiis/\fir~$<0.33\times10^{-3}$, consistent with local ULIRGs and
lower than most detected local star forming galaxies. Similarly to the
\oi\ data (Section~\ref{sec:oi}), the dichotomy between the individual
NGP.NA.144 detection and the stacked \oiii\ measurement is suggestive
of a range of conditions in the \oiii\ emitting region of SMGs.

\subsubsection{\niii\micron}
\label{sec:niii}

\niii\ has similar ionization energy and critical density as \oiii\
and is also a strong tracer of \hii\ regions. Local ULIRGs have a
similar \niii\ deficit to \oiii\ and other fine structure lines.  We
targeted \niii\ in four lensed SMGs, and none were individually
detected above $3\sigma$, although the \niii\ line is detected in the
stack of eight SMGs at $3.0\sigma$.  The stack, as well as the four
individual targets, have \niiis/\fir$_{(42.5-122.5\,\micron)}$ values that are broadly
consistent with local sub-LIRGs, although still within the upper range
of values of local ULIRGs. Overall, the \niii\ emission regions in
SMGs likely have similar conditions to local star-forming galaxies and
some local ULIRGs.

\subsubsection{\oiv\micron}
\label{sec:oiv}

\oiv\ is a high excitation line, and as such is a reliable
tracer of AGN activity \citep[e.g.,][]{Melendez08, Rigby09}. As can be seen in Figure~\ref{fig:ratios}
all local sources with \oivs/\fir$_{(42.5-122.5\,\micron)}\gtrsim10^{-3}$ are AGN. Other galaxy types
can have some  \oiv\ emission, but it is always fainter (relative to
\fir) than this limit, since the \oiv\ contribution is enhanced in AGN environments.

Of our 10  \oiv\ targets, nine are undetected and one is
detected at  $3.1\sigma$ (HXMM01). All of the undetected
systems have upper limits of \oivs/\fir$\simeq2\times10^{-2}$, and therefore we cannot rule
out some AGN contribution to these galaxies. For HXMM01 we measure
\oivs/\fir$=(3.9\pm1.3)\times10^{-3}$, which is significantly above the observed ratios in
local star-forming galaxies, and an indication of a hidden AGN
in this system. HXMM01 was previously studied by \citet{Fu13}, who
found no evidence of an AGN in the mid-IR (IRAC) or far-IR colours, or
in (shallow) X-ray observations. They did observe a broad H$\alpha$
line in HXMM01, but concluded that it was likely driven by starburst
outflows. Thus, our \oiv\ observation is currently the only distinct
evidence of an AGN in HXMM01. 

\oiv\ is not detected in the stacked spectrum, placing an upper limit
on \oivs/\fir\ of $1.4\times10^{-3}$ in average SMGs -- within the
region inhabited by local AGN, star-forming galaxies and LINERs. Thus,
there is no evidence from the \oiv\ that typical SMGs contain strong AGN, but
we cannot rule out lower luminosity AGN with the current data.  This
is consistent with previous analyses of SMGs using other AGN tracers
\citep[e.g.,][]{Alexander05, Pope08, Wang13}, and broadly consistent
with the results from \sixii\ and \sxiii\ (Section~\ref{sec:agnsis}),
hinting at some AGN emission if the AGN in SMGs are weak and therefore
not picked up by our relatively shallow \oiv\ data.

\subsubsection{\feii\micron}

\feii\ was not specifically targeted by this program, as it is rarely
detectable, even locally. However, it occurs only  $\sim0.1$\,\micron\
(rest-frame) away from \oiv\ and is therefore included in the spectral
coverage of the 10 SMGs for which we targeted the \oiv\ transition.
\feii\ is predominantly emitted from PDRs and it has a low ionization
energy and high critical density, similar to \sixii. Also similarly
to \sixii, \feiis/\fir$_{(42.5-122.5\,\micron)}$ is weakly anti-correlated with \lfir\ (40--500\,\micron) locally
(Figure~\ref{fig:ratios}), although \feii\ has been measured in fewer
galaxies for comparison \citep[e.g.,][]{Verma03, Farrah07, OHalloran08,
  BernardSalas09}. As our \feii\ data were also obtained serendipitously, they are
shallower than required to probe the local observed range of
\feiis/\fir. All the individual targets as well as the stack are
undetected, with limits consistent with local observations.

\begin{figure*}
\centering
\begin{tabular}{cc}
\includegraphics[width=8.4cm]{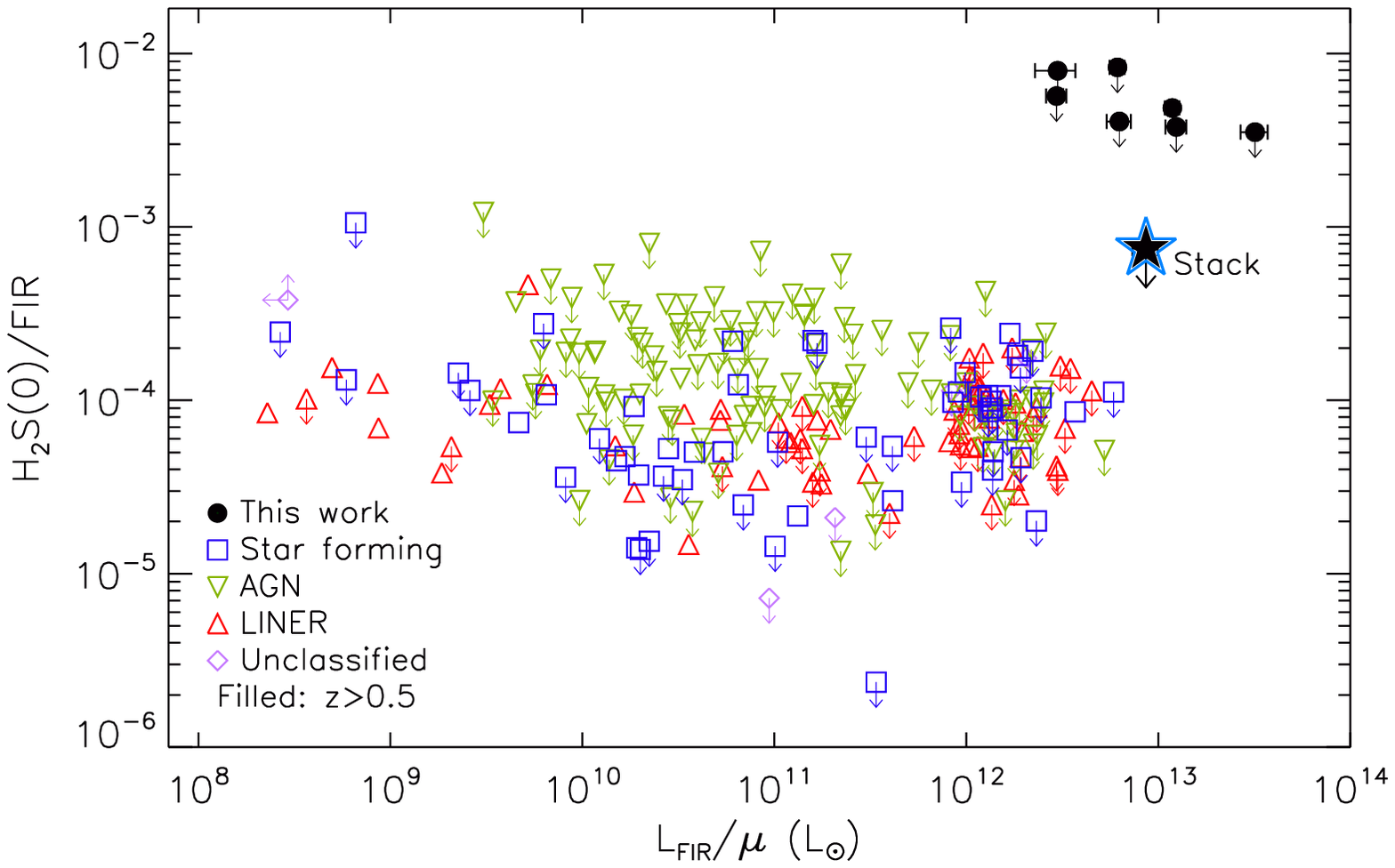} & 
\includegraphics[width=8.4cm]{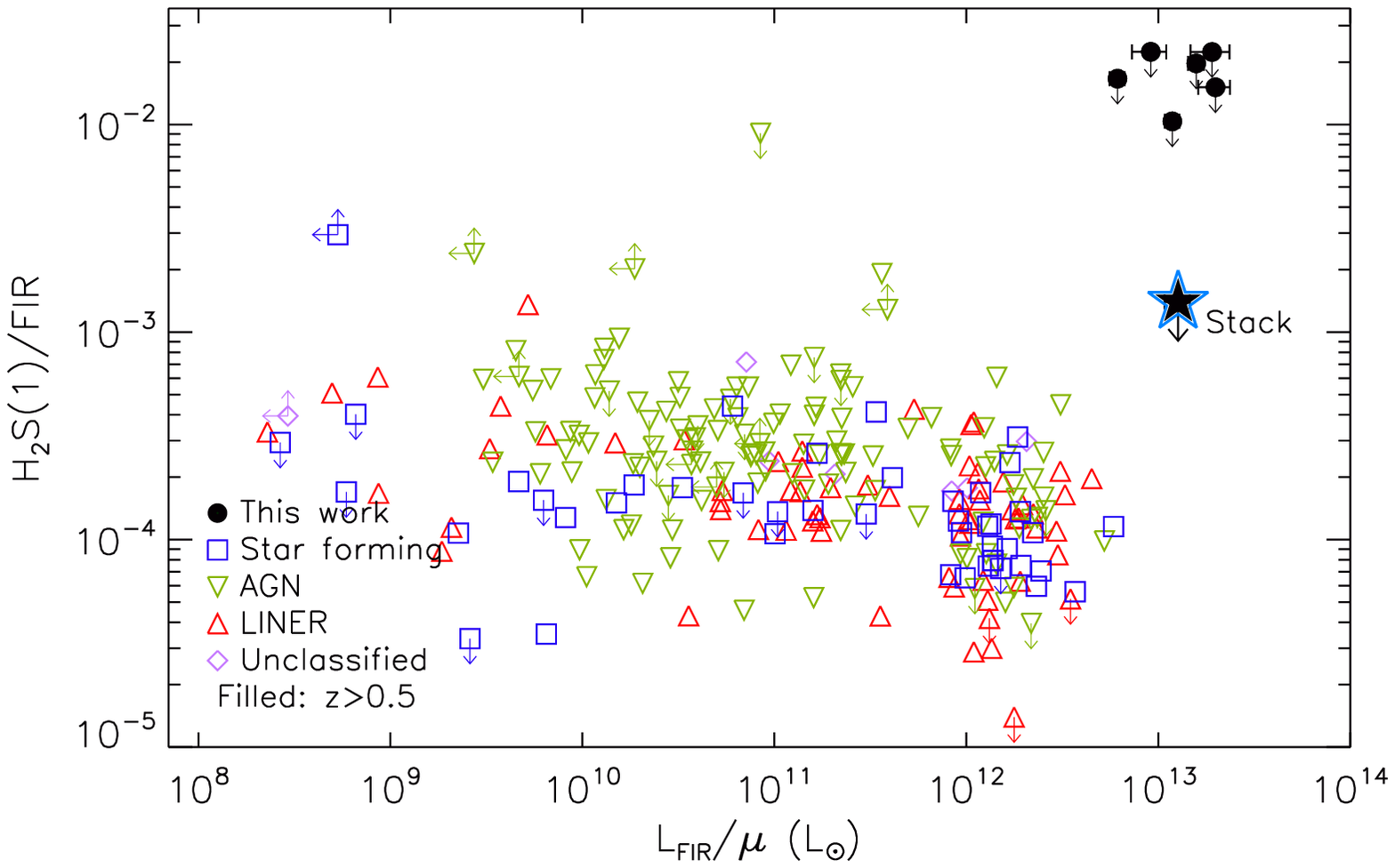}
\end{tabular}
\caption{As Figure~\ref{fig:ratios} for the H$_2$ rotational transitions.}
\label{fig:h2ratios}
\end{figure*}

\subsubsection{\htsz\ and \htso}

The \htsz\ and \htso\ rotational transitions trace warm
($T\simeq150$\,K) gas reservoirs \citep[e.g.,][]{Rigopoulou02, Roussel07,
  Nesvadba10, Higdon14}, winds \citep[e.g.,][]{Beirao15}, and likely
shocked gas in extreme systems with enhanced \htwo\ emission
\citep[e.g.,][]{Appleton06}.  Thus, with detections of \htsz\ and
\htso\ we would be able to directly measure the warm \htwo\ gas mass
(rather than purely relying on CO), and use the prevalence of shocks
to identify violent merger activity. However, the data are
not as deep as expected (Figure~\ref{fig:h2ratios}), and none 
of the individual SMGs nor the stack have emission detectable at the
$\ge3\sigma$ level.

\begin{deluxetable}{lc}
\centering
\tablecaption{Relative line strengths for average SMGs. 
  \label{tab:stackratio}} 
\startdata 
\hline  
Ratio & Value \\
\hline\hline
\multicolumn{2}{c}{Measurements$^a$}\\
\oi/\fir\ & $(3.6\pm1.2)\times10^{-4}$ \\
\sxiii/\fir\ & $<3.6\times10^{-4}$ \\
\sixii/\fir\ & $(8.4\pm1.7)\times10^{-4}$ \\
\oiii/\fir\ & $<3.3\times10^{-4}$ \\
\niii/\fir\ & $(2.7\pm1.0)\times10^{-4}$ \\
\oiv/\fir\ & $<1.4\times10^{-3}$ \\
\feii/\fir\ & $<1.4\times10^{-3}$ \\
${\rm H}_2$ S(0)/\fir\ & $<7.5\times10^{-4}$ \\
${\rm H}_2$ S(1)/\fir\ & $<1.4\times10^{-3}$ \\
\cii/\fir\ & $(1.7\pm1.1)\times10^{-3}$ \\
\oi/\cii & $2.2\pm1.5$\\
\hline\hline
\multicolumn{2}{c}{Predictions from the PDR model$^b$} \\
{[\ion{C}{1}]}609/\fir\ & (0.01--20)$\times10^{-5}$\\
{[\ion{C}{1}]}370/\fir\ & (0.01--60)$\times10^{-5}$\\
{[\ion{O}{1}]}145/\fir\ & (0.01--20)$\times10^{-5}$\\
\feii/\fir, $Z=Z_{\odot}$ & (0.3--50)$\times10^{-7}$\\
\feii/\fir, $Z=3Z_{\odot}$ & (0.9--800)$\times10^{-7}$
\enddata
\tablecomments{
Upper limits are $3\sigma$ limits. \fir\ refers to the
42.5--122.5\,\micron\ continuum flux (Section~\ref{sec:seds}). 
$^a$From the mean stack measurements (Section~\ref{sec:linestack}).
$^b$As discussed in Section~\ref{sec:pdrparam}, we use the best-fit
parameters of the PDR model to predict the average strengths of other
transitions from the PDRs of the SMGs.}
\end{deluxetable}

\subsection{AGN contribution traced by \sixii\micron\ and
  \sxiii\micron\ emission}
 \label{sec:agnsis}

Previous studies have shown that the ratio of \sixii\ to \sxiii\ flux
is an effective discriminator between AGN, LINERs and \hii\ regions
\citep{Dale06, Dale09}. Typically the \sixii/\sxiii\ ratio
is used in conjunction with a second discriminator such as the ratios
of [\ion{Ne}{3}]~16\micron\ to [\ion{Ne}{2}]~13\micron\
\citep{Dale06} or
[\ion{Fe}{2}]~26\micron\ to [\ion{Ne}{2}]~13\micron\
\citep{Dale09}. Observations of these additional lines are not
available for SMGs, but the \sixii/\sxiii\ ratio alone is still useful.

Five of our individual targets and the stack have data for both the \sixii\ and
\sxiii\ transitions. However, none have $\ge3\sigma$ detections in one or
both of these lines, and therefore the AGN contribution cannot be
traced on a individual target basis from our \sixii\ and \sxiii\ data. 
For the stacked data, the \sixii\ detection and the \sxiii\ limit give
\sixii/\sxiii~$\ge3.35$ ($3\sigma$), placing the average SMG in region I
and II of \citet{Dale06}, corresponding to AGN and LINER emission.
Thus, the \sixii\ and \sxiii\ measurements suggest that SMGs {\it on
  average} contain AGN, although the absence of other AGN tracers in
most cases \citep[e.g.,][]{Alexander05, Pope08, Laird10, Wang13}
indicate that these are unlikely to dominate the energetics.

\begin{figure}
\centering
\includegraphics[width=8.4cm]{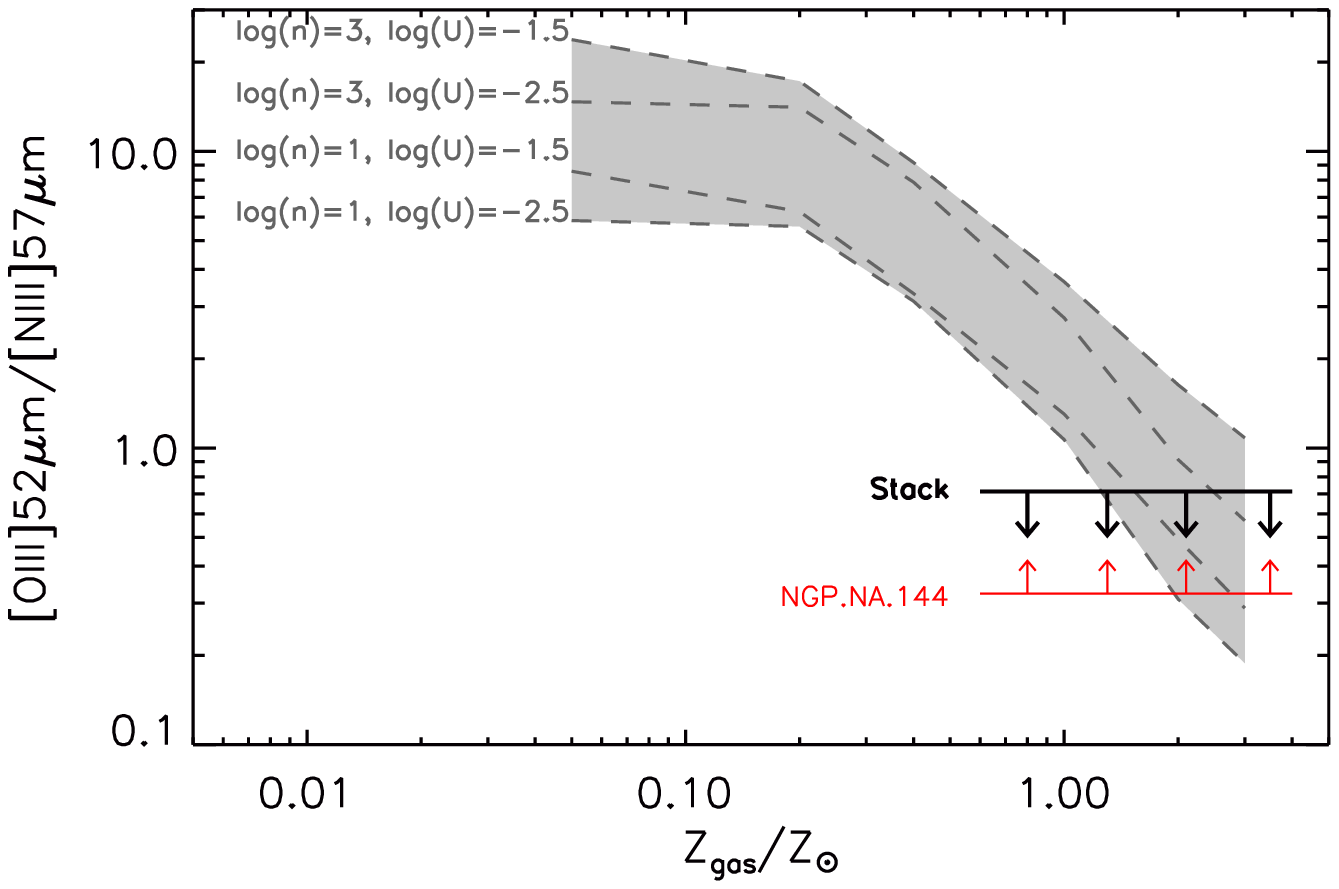}
\caption{\oiii/\niii\ ratio as a function of gas-phase
  metallicity. The measured ratio for the stacked SMGs and NGP.NA.144
  -- the only individual target with a $\ge3\sigma$ detection in at
  least one of the relevant transitions -- are shown.  The shaded
  region and dotted lines show the ratio predicted by \citet{Nagao11}
  for systems with different gas density ($n$ in ${\rm cm^{-3}}$ and
  dimensionless ionization parameter, $U$). The data for NGP.NA.144 are
  not constraining but the low limit on the \oiii/\niii\ ratio in the
  stack indicates that SMGs are enriched to $Z\gtrsim Z_{\sun}$ on
  average.  }
\label{fig:metals}
\end{figure}

\subsection{Metallicity from the \oiii\micron\ and \niii\micron\ emission}
\label{sec:metals}

As shown by \citet{Nagao11} the \oiii/\niii\ flux ratio can be a
tracer of gas-phase metallicity (see also Pereira-Santaella et al.\ in
prep.). In Figure~\ref{fig:metals} we compare
the measured \oiii/\niii\ ratio for the stack and for NGP.NA.144,
which is the only individual SMG with a detection in at least one of
the relevant transitions.  

The models in Figure~\ref{fig:metals} are from \citet{Nagao11} and
show the variation of \oiii/\niii\ with metallicity for different
ionization parameters (${\rm log_{10}}(U)=-2.5$ to $-1.5$;
dimensionless) and gas densities (${\rm log_{10}}(n/{\rm cm^{-3}})=1$
to 3), compared with the measurements for SMGs.  These models are generated
with {\sc cloudy} \citep{Ferland98} and include PDRs and \hii\
regions, although \citet{Nagao11} show that the \oiii\ and \niii\
emission is mostly from the \hii\ regions.  The range of densities
investigated by \citet{Nagao11} is consistent with the values that we
find for the PDRs in average SMGs (Section~\ref{sec:model}). The
ionization parameter ($U$) used by \citet{Nagao11} to trace the
strength of the ionizing source is defined as the ratio of
hydrogen-ionizing photons to total hydrogen density. They consider
values of $U$ that are typical of \hii\ regions -- the main sources of
\oiii\ and \niii\ emission -- and are therefore valid for our SMGs. 
These model $U$ cannot be directly compared with the $G_0$ from our PDR results
(Section~\ref{sec:model}) because $U$ is dependant on
total hydrogen density, whereas the hydrogen in
PDRs is primarily atomic.\footnote{It is pertinent to note that
  PDR analyses are usually most sensitive to $G_0/n$, which similarly
  to $U$ is a ratio of photon to gas density, although $U$ and $G_0/n$
  trace different phases of material.}

Metallicity measurements from \oiiis/\niiis\ are not expected to be
significantly affected by different optical thickness of the \oiii\
and \niii\ lines, because the wavelength difference is small. \oiii\
and \niii\ also have similar filling factors and the ratio is not
affected by differential magnification since the two lines have
similar ionization parameters and critical densities to each other,
and are therefore emitted from approximately the same region of the galaxies. There
is unlikely to be a major effect from any weak AGN
component, since neither line is significantly boosted by AGN
emission. However, the presence of AGN could enhance the ionization
parameter, which, as can be seen from Figure~\ref{fig:metals} would
serve to increase the metallicity for a given observed
\oiii/\niii. However, if
AGN emission is dominant (unlikely for SMGs; e.g.\
Section~\ref{sec:agnsis}), the \oiiis/\niiis\ ratio is no longer a
good metallicity tracer, since the models are for \hii\ regions and
not XDRs.

The results from the stacked data (Figure~\ref{fig:metals}) show that
the average  \oiii/\niii\ ratio of SMGs is indicative of them
containing enriched gas, with average metallicities,
$Z\gtrsim Z_{\sun}$. The \oiii\ and \niii\ data are insufficient to
constrain the metallicity of NGP.NA.144. 
Previous measurements of the metallicities of SMGs are
 hard to come by and have large uncertainties. There have been indications that they typically have
 approximately sub-solar to a few times solar metallicities -- including from the
 [\ion{N}{2}]$\lambda6584$\AA/H$\alpha$ \citep{Swinbank04}, extrapolations
 from the mass-metallicity (or mass-metallicity-SFR) relation, and
 [\ion{N}{2}]205\micron/\cii\micron\ for one SMG at $z\sim4.8$
 \citep{Nagao12} -- although these measurements were plagued with
 uncertainties and systematic effects.

\subsection{ISM density and FUV radiation field }
\label{sec:model}

Rather than considering each line observation in isolation, more can
be learned by examining the emission line ratios in concert with PDR
modelling. However, the low signal-to-noise ratios of the individual observations
means that this is only possible for the stacked data -- i.e. we can
only examine the properties of the average SMG. 

\begin{figure*}
\centering
\includegraphics[width=8.4cm]{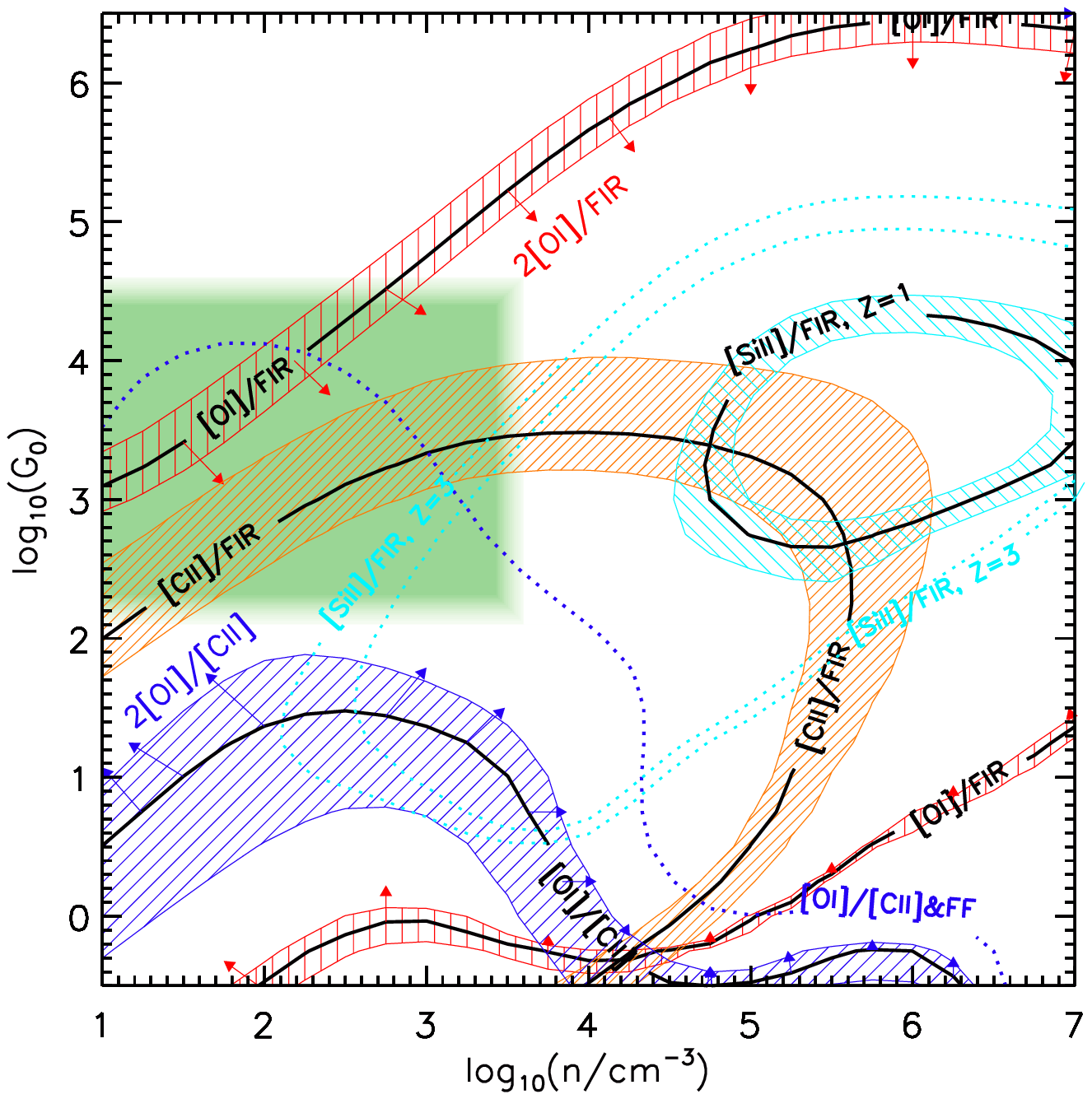}
\includegraphics[width=8.4cm]{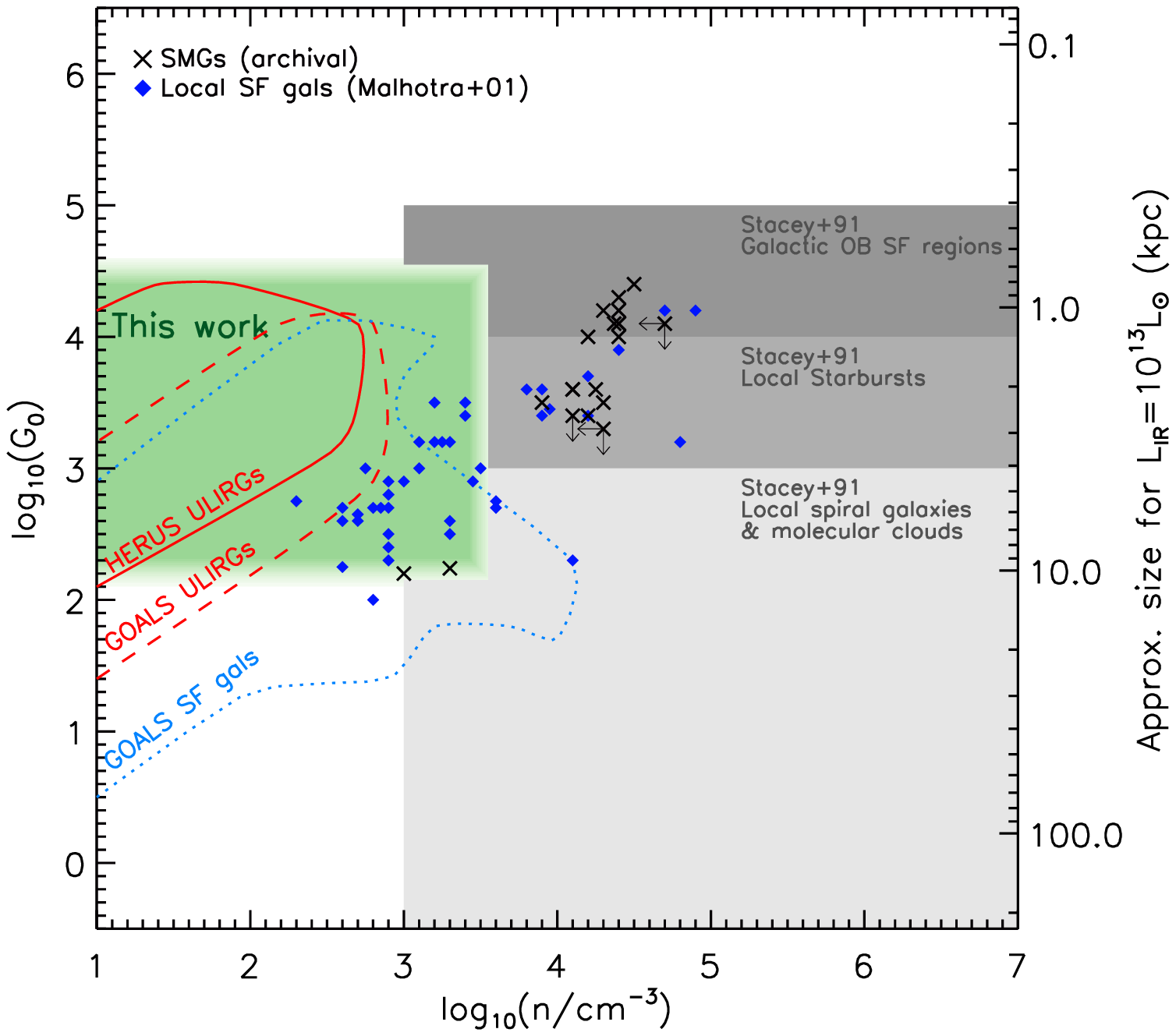}
\caption{{\it Left:} Contours showing constraints on the gas density ($n$) and
  FUV field strength ($G_0$ in units of the Habing Field) for the
  \citet{Kaufman99, Kaufman06} PDR model from our mean stacked
  spectra. Thick black lines represent the average of the different
  line ratios (as labeled), and the colored regions represent the
  $\pm1\sigma$ measurement uncertainties. \citet{Kaufman06} includes
  two models for \sixii\ with metallicity equal to the solar
  neighbourhood value ($Z=1$) and for three times solar ($Z=3$), with both shown. There is also the possibility
  that the \oi\ emission is self-absorbed and so we include vectors on
  the \ois/\fir\ and \ois/\ciis\ contours to show how the positions of
  the contours would change if the average intrinsic \oi\  strength is twice
  that measured (or for \ois/\ciis\, the measured \cii\ is twice that
  from the PDR alone; Section~\ref{sec:caveat}). The dotted blue line
  labeled ``\ois/\ciis\&FF'' shows where the \ois/\ciis\ contour would lie
  if the measured value were corrected according to the relative
  filling factors of \ois\ and \ciis\ from M82
  (section~\ref{sec:caveat}). The green shaded region represents the
  region of $n$--$G_0$ space that we determine is most representative
  of the average SMG, once the additional considerations described in
  Section~\ref{sec:caveat} are taken into account
  (Section~\ref{sec:pdrparam}). 
  {\it Right:} Comparison of the gas density and FUV field strength
  for average SMGs (derived here; green shaded) compared with local
  star-forming galaxies \citep{Malhotra01} and existing SMG
  measurements. The existing archival SMG measurements are mostly
  derived from only \cii\ and CO data \citep{Sturm10, Cox11,
    Danielson11, Valtchanov11, AlaghbandZadeh13, Huynh14, Rawle14} and
  have typical uncertainties of 0.2--0.5 in both
  ${\rm log}_{10}(n/{\rm cm^{-3}})$ and ${\rm log}_{10}(G_0)$. The
  right-hand axis shows approximate sources sizes for a SMG with
  $L_{\rm IR}=10^{13}$\lsun\ at different $G_0$ values
  (Section~\ref{sec:pdrparam}).  Contours highlight the regions that
  typical ULIRGs inhabit, from the GOALS (D\'iaz-Santos et al. in
  prep.) and HERUS samples \citep{Farrah13}, as well as GOALS
  star-forming galaxies (D\'iaz-Santos et al. in prep.).  We also show
  the areas that are found to be dominated by local starbursts, local
  spiral galaxies and Galactic molecular clouds, and Galactic OB
  star-forming regions on the basis of \cii/\aco\ measurements
  \citep{Stacey91}.  The density and FUV field strength from our
  fine-structure line data are similar to individual local
  star-forming galaxies but lower in density than the majority of
  individually measured SMGs (discussed in
  Section~\ref{sec:pdrparam}). The comparison with GOALS and HERUS
  shows that we do not have sufficient data to robustly distinguish
  whether the SMGs have internal conditions more similar to local
  ULIRGs or star-forming galaxies.  }
\label{fig:pdrcont}
\end{figure*}

We use the PDR models of \citet{Kaufman99, Kaufman06}, accessed via
{\sc PDR Toolbox}\footnote{\url{http://dustem.astro.umd.edu/pdrt}}
\citep{Pound08}. The model is characterized using a varying gas
density ($n$, in units of the density of hydrogen nuclei), and the
strength of the FUV (energies $h\nu=6$--13\,eV) radiation field
($G_0$, in units of the Habing Field,
$1.6\times10^{-3}{\rm erg\,cm^{-2}\,s^{-1}}$).
Figure~\ref{fig:pdrcont} highlights where the model $n$ and $G_0$
produce \ois/\fir$_{(42.5-122.5\,\micron)}$ and
\sixiis/\fir$_{(42.5-122.5\,\micron)}$ consistent with the SMG average
values measured from the fiducial mean stacks
(Section~\ref{sec:linestack} and Table~\ref{tab:stackratio}). \feii\
is also available in the \citet{Kaufman06} PDR model, but our
non-detection is too shallow to be useful in constraining $n$ and
$G_0$, and so it is not included in Figure~\ref{fig:pdrcont} and the
following discussions.

The \oi\ and \sixii\ data alone cannot constrain the conditions of the
PDRs in SMGs, so we also include archival \cii\ measurements from
\citet{Gullberg15} and \citet{George15} for gravitationally lensed
SMGs. Many of the sources in these papers were included in our PACS
stacks. The mean \ciis/\fir$_{(42.5-122.5\,\micron)}$ and \ois/\ciis\
are shown in Figure~\ref{fig:pdrcont}, where both the \cii\ and \oi\
fluxes are scaled by the \fir\ (42.5-122.5\,\micron) of the sources
measured, so as to remove any luminosity effects.

To interpret ISM conditions via spectroscopy and PDR models it is
typical to identify the regions of $n$ and $G_0$ space where the
constraints from different line measurements overlap
\citep[e.g.,][]{Kaufman99, Kaufman06, Pound08}. It is important to note
that the PDR models assume that all the measured fluxes are being
emitted from the same spatial region. However, for our SMGs PACS
cannot resolve different regions and thus each line measurement is an
aggregate over the whole galaxy. Since different PDRs within a galaxy
may have different properties, the different line observations may,
therefore, be differently weighted towards different regions.
Furthermore, we are investigating stacked data -- i.e.\ average line
strengths over many SMGs -- and therefore emission from
several galaxies, which may also have intrinsic spread in their
properties (Section~\ref{sec:ratios}). Therefore, our conclusions are
averages, with some natural weighting towards more line-luminous
regions and galaxies. In Section~\ref{sec:caveat} we discuss
 further considerations; our final interpretation of the PDR
parameters are discussed in Section~\ref{sec:pdrparam}.

\subsubsection{Additional considerations}
\label{sec:caveat}

There are several factors that affect the interpretation of
Figure~\ref{fig:pdrcont}, which we now discuss.
Firstly, a small fraction of local galaxies exhibit self-absorption in
the \oi\ line \citep[e.g.,][]{Fischer97, Fischer99, Genzel00, Farrah13, Rosenberg15}, and it is
possible that the  \oi\ emission of SMGs may be self-absorbed since our
data are too shallow to directly measure this via the shape of the line. Therefore, in
Figure~\ref{fig:pdrcont} we demonstrate that a factor of two
increase in the \oi\ flux from the measured value would have only a
minor effect on the positioning of the \ois/\fir$_{(42.5-122.5\,\micron)}$ and \ois/\ciis\
contours, with both being within $2\sigma$ of the directly measured
values. The factor of two shown demonstrates the approximate maximum
change in Figure~\ref{fig:pdrcont} likely from \oi\
self-absorption. 

\begin{figure}
\includegraphics[width=8.4cm]{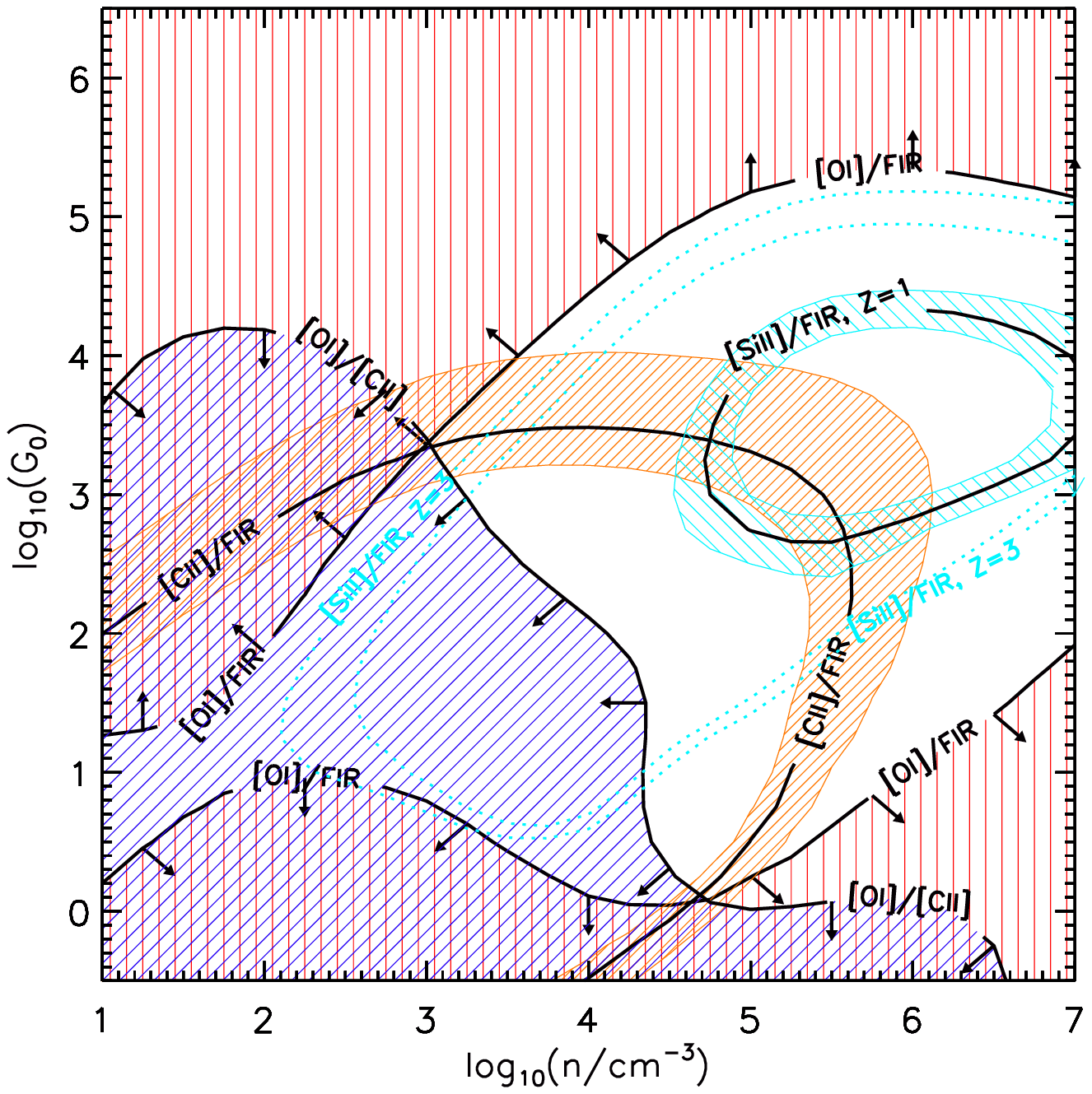}
\caption{Contours showing the effect of the constraints on the gas density ($n$) and
  FUV field strength ($G_0$ in units of the Habing Field) for the
  \citet{Kaufman99, Kaufman06} PDR model if the $3\sigma$ limit on
  \oi\ emission from the \fir-weighted stacked
  is used rather than the fiducial mean stack (shown in
  Figure~\ref{fig:pdrcont}). Colours are as  Figure~\ref{fig:pdrcont}
  with shading showing acceptable regions based on each stacked
  measurement, with the addition of black arrows to highlight that
  both the \ois/\fir\ and \ois/\ciis\ are $3\sigma$ upper limits. }
\label{fig:pdrweighted}
\end{figure}

Secondly, as discussed in Section~\ref{sec:linestack} there is a
difference in the \oi\ emission as measured from the (fiducial) mean
stack, and the \fir-weighted mean stack. If we use the limit on \oi\
from the weighted mean stack (instead of the fiducial mean stack;
Figure~\ref{fig:pdrcont}) the acceptable
\ois/\fir$_{(42.5-122.5\,\micron)}$ and \ois/\ciis\ regions would both
be substantially larger than the fiducial result, as shown in
Figure~\ref{fig:pdrweighted}.  Overall, considering the weighted stacked
fluxes expands the acceptable $G_0$ range (before accounting for size
arguments; Section~\ref{sec:pdrparam}), but has minimal effect on the
inferred PDR density.

In addition, the \citet{Kaufman06} model includes two sets of \sixii\
data, for two different metallicities -- labelled $Z=1$ (where the gas phase
metallicities are those in the solar neighbourhood) and $Z=3$ (where
all elements are three times more abundant) in Figure~\ref{fig:pdrcont}. As discussed in
Section~\ref{sec:metals} SMGs are likely to have $Z\gtrsim Z_{\sun}$,
although we cannot distinguish between $Z=Z_{\sun}$ and $Z=3Z_{\sun}$
with current data. Due to the uncertainties we include the constraints
for both model metallicities in Figure~\ref{fig:pdrcont}. 

For the main contours on Figure~\ref{fig:pdrcont} the relative filling factors of the various line
species in the PACS beams are considered to be equal, i.e.\ we have
not applied any corrections for filling factors.  Such corrections are
expected to have the biggest effect on the \ois/\ciis\ ratio.
 To make corrections due to the relative \oi\ and \cii\
filling factors we would need to know the relative sizes of the
regions that dominate those emission lines. The large PACS (for \oi) and
APEX or SPIRE (for \cii) beam sizes preclude directly measuring the
extent of the emission. Instead, to gain some insight into the
possible size of this effect, we consider the local starburst M82,
where the extents of the \oi\ and \cii\ emission regions can be
directly measured, leading to a required correction factor of
0.112 on the \cii\ flux, i.e.\ the \ois/\ciis\ is increased by a
factor of $1/0.112=8.9$ \citep[e.g.,][]{Stacey91, Lord96, Kaufman99, Contursi13}. Under these circumstances
the low density ($n\lesssim10^3\,\rm{cm^{-3}}$) end of the \ois/\ciis\
contour on Figure~\ref{fig:pdrcont} would be shifted to higher $G_0$
(demonstrated with the dotted line on Figure~\ref{fig:pdrcont}), with
the $1\sigma$ uncertainty region encompassing $G_0=10^{1-6.5}$,
$10^{2-5.5}$ and $10^{1.8-4.5}$ for $n=10^{1, 2, 3}\,\rm{cm^{-3}}$,
respectively. The M82  filling
factor correction is substantial, and thus the correction to the 
\ois/\ciis\ ratio for the average
SMG is likely to be lower than this; thus the correction explored
here demonstrates an approximate upper boundary to the size of the
effect.

Another related consideration is that some of the line emission may originate
from \hii\ regions (or other gas)  rather than  PDRs. This is most
likely to affect the \cii\ flux due to the critical densities and
ionization parameters of the different transitions studied here, and
therefore the non-PDR \cii\ emission should be subtracted from the
observed \cii\ intensity prior to using it to analyse the PDR conditions. This is
typically done by using multi-phase modelling (e.g. {\sc cloudy};
\citealt{Ferland98}), or using other transitions (such as \nii\micron)
to determine the contribution from \hii\ regions. However, there are
few observations of \nii\ in SMGs  \citep{Ferkinhoff11, Combes12,
  Decarli12, Nagao12}, and the
complexity of {\sc cloudy} modelling coupled with the limitations in our data
means that {\sc cloudy} analysis will not improve the uncertainties in
our analysis.  We instead investigate the effect of some of the
\cii\ emission coming from non-PDR gas qualitatively, noting that if
some of the observed \cii\ flux is not from the PDRs, then the correct
values of \ciis/\fir$_{(42.5-122.5\,\micron)}$ and \ois/\ciis\ to use in Figure~\ref{fig:pdrcont}
would be decreased and increased, respectively. The effect of a factor
of two increase in \ois/\ciis\ is shown on Figure~\ref{fig:pdrcont}
and is minor. Note that due to the critical density and ionization potential of
\cii\ it is unlikely to be dominated by non-PDR emission, i.e., the
factor of two considered is the approximately the upper limit of any
non-PDR correction required. A decrease of \ciis/\fir$_{(42.5-122.5\,\micron)}$ also has a small effect on
Figure~\ref{fig:pdrcont}, shifting the contours to slightly higher
$G_0$ and $n$ but remaining within the current $1\sigma$ uncertainty
area, although the updated error region marginally overlaps with the
\ois/\fir$_{(42.5-122.5\,\micron)}$ uncertainties. Thus different non-PDR emission is unlikely to
have a substantial effect on the interpretation of the \cii\ and \oi\
fluxes in Figure~\ref{fig:pdrcont}.

In some cases \sixii\ can be boosted by AGN emission
\citep[e.g.,][]{Dale06, Dale09}. If some of the observed \sixii\ flux
is from AGN then removing this contribution would move the \sixiis/\fir$_{(42.5-122.5\,\micron)}$
contours on Figure~\ref{fig:pdrcont} outwards, i.e.\ similarly to
increasing metallicity. As discussed in Section~\ref{sec:agnsis} there
is evidence from the \sixiis/\sxiiis\ ratio that there is some AGN
contribution in the stacked \sixii\ data, which may explain why the contours for $n$ and
$G_0$ for our observed \sixiis/\fir$_{(42.5-122.5\,\micron)}$ (particularly for $Z=Z_{\sun}$) are offset from those derived from
\ois/\fir$_{(42.5-122.5\,\micron)}$ and \ois/\ciis.

It is possible that the different transitions studied may have
different optical depths. Although all the lines in
Figure~\ref{fig:pdrcont} are in the IR they cover a significant range
in rest-frame wavelength (34\micron\ for \sixiis, to 158\micron\ for
\ciis) and we are examining some of the dustiest galaxies in the
Universe. Similarly to the filling factors, this effect is most likely
to affect the positioning of the \ois/\ciis\ contours, and will shift
them in a similar manner (i.e. towards higher $G_0$ for a given $n$),
due to the shorter wavelength \oi\ being more strongly affected.

 Finally, $\sim60\%$ of the galaxies included in the stacks are known
 to be gravitationally lensed, and we have so far assumed that the
 lensing amplification is equal in all components of emission. In
 fact, since these are composed of galaxy-galaxy lenses, differential
 magnification, caused by different regions of the background galaxy
 being amplified by different amounts is possible. If differential
 magnification is a random effect then it is more likely to affect analyses of
 individual galaxies (e.g. in Figure~\ref{fig:ratios}) than the
 average values examined in Figure~\ref{fig:pdrcont} and for the PDR
 modelling, where the effects will be minimized due to averaging
 many sources. However,  there may be systematic effects in regions
 emitting the majority of the different line species, resulting in
 them being differentially amplified. This is likely to be an
 important effect, due to biases in the
 identification of lensed SMGs -- a crucial step in the selection of
 many of the PACS targets.

 The spatial resolution of the spectroscopy is insufficient to resolve
 and model the lensing for  each transition individually, so even if we had
 attained detections for several galaxies individually we would be
 unable to determine the differential magnification on a case by case
 basis. Instead we consider the simulations
 of \citet{Serjeant12} who investigated systematic effects in the
 differential magnification of a simple dusty galaxy model with a
 variety of foreground galaxy lenses and alignments.  \citet{Serjeant12}
 found no systematic differential magnification effects in \ciis/\fir$_{(42.5-122.5\,\micron)}$
 for lensed SMGs and claimed that that there are similarly unlikely to
 be systematic effects in \ois/\fir$_{(42.5-122.5\,\micron)}$ because \oi\ and \cii\ are
 observed to be co-spatial in M82 on small scales \citep{Sturm10}. In
 that scenario we would also not expect differential amplification
 effects in \ois/\ciis, or \sixiis/\fir$_{(42.5-122.5\,\micron)}$, which traces the same
 environments.  However, we consider how the contours on
 Figure~\ref{fig:pdrcont} would change if half the \cii\ emission is
 instead from a more extended region (e.g. an \hii\ region) than the
 PDRs. In this case the \cii\ flux from the smaller PDR will typically
 be more highly magnified than the extended \cii\ and thus the overall
 effect will be to minimize the fraction of detected \cii\ from
 non-PDR regions, somewhat canceling out the effect of non-PDR \cii\
 emission on the line ratios. In this case the \sixii\ and \oi\
 emitting gas will be situated in the PDRs along with the \cii, and thus be
 co-spatial on all but the smallest scales, making them unlikely to be
 strongly systematically affected by differential magnification. We
 reiterate that this discussion of the
 systematic effects of differential amplification is applicable only
 to the average (i.e. stacked) values of the sample, and that
 individual galaxies may have quite substantial differential
 magnification effects.

\subsubsection{Inferred PDR parameters}
\label{sec:pdrparam}

Based on the data shown in Figure~\ref{fig:pdrcont} and the discussion
in Section~\ref{sec:caveat} we determine that the average SMG has
$n\sim10^{1-3.5}{\rm cm^{-3}}$ and $G_0\sim10^{2.2-4.5}$.  In
Table~\ref{tab:stackratio} we use these values and {\sc PDR Toolbox}
to predict the average strength of the [\ion{C}{1}]370\micron,
[\ion{C}{1}]609\micron, [\ion{O}{1}]145\micron, and \feii\micron\
lines from the PDRs in SMGs.

The $n\sim10^4\,{\rm cm^{-3}}$ and $G_0\sim10^{-0.2}$ solution where the
\ois/\fir$_{(42.5-122.5\,\micron)}$, \ciis/\fir$_{(42.5-122.5\,\micron)}$, and \ois/\ciis\ contours are coincident
(Figure~\ref{fig:pdrcont}) is excluded because such a low $G_0$ in
SMGs with (intrinsic) $L_{\rm IR}\sim10^{12.5-13}$\lsun\ would require
source sizes of hundreds of kpc, which is clearly unphysical.\footnote{
  \citet{Wolfire90} showed that $G_0 \propto L_{\rm IR}/R^2$, where $R$ is
  a characteristic size of the emission region. The proportionality
  constant includes the contribution from the ionizing photon field
  (i.e., dependant on the IMF and star-formation history); we use the
  values from \citet{Stacey10} to estimate the sizes described here
  and shown in Figure~\ref{fig:pdrcont} \citep[see
  also][]{Danielson11}.} For $G_0\sim10^{2.2-4.5}$ and
$L_{\rm IR}\sim10^{13}$\lsun\ the source sizes are expected to be
$\sim1$--10~kpc; lower $G_0$ values are excluded as they would require
excessively large sources. Measurements show that SMGs have typical
total sizes of $\sim 0.5$--10~kpc \citep{Tacconi06, Younger08,
  Swinbank10, Ivison11, Riechers11, Bussmann13, Calanog14,
  Ikarashi15}, usually smaller in the far-IR continuum (dust) than the
rest-frame optical (stars) or radio emission
\citep[e.g.,][]{Simpson15}. Higher $G_0$ values are acceptable if the
weighted stack for the \oi\ line is used, although for $G_0\gtrsim10^{5}$ the
inferred source sizes would be smaller than typically observed for
SMGs and are thus unlikely. 

In the right-hand panel of Figure~\ref{fig:pdrcont} the values of $n$
and $G_0$ that we infer for average SMGs are compared with the regions
of $n$--$G_0$ space typically populated by local ULIRGs and
star-forming galaxies, as determined by similar PDR modeling from
HERUS \citep{Farrah13} and GOALS (D\'iaz-Santos et al.\ in prep.). The
HERUS and GOALS ULIRGs include galaxies with
$L_{\rm IR}\ge10^{12}$\,\lsun (with $>95\%$ star-formation rather than
AGN dominated); the GOALS star-forming sample are the galaxies with
$>50\%$ contribution to the bolometric luminosity from star-formation
of which $\sim 90\%$ are LIRGs and the remainder ULIRGs.  Also
highlighted in Figure~\ref{fig:pdrcont} are the regions found to be
preferentially occupied by local starbursts, spiral galaxies,
molecular clouds and galactic OB star-forming regions, as determined
by \cii/\aco\ ratios \citep{Stacey91}.  In addition, we also show
measurements of local star-forming galaxies \citep{Malhotra01} and
individual SMGs with existing measurements of $n$ and $G_0$, typically
from CO and \cii\ lines \citep{Sturm10, Cox11, Danielson11,
  Valtchanov11, AlaghbandZadeh13, Huynh14, Rawle14}.

It can be seen in Figure~\ref{fig:pdrcont} that full PDR modeling
\citep[from the HERUS and GOALS results; ][D\'iaz-Santos et al.\
in prep.]{Farrah13} results in correlated $G_0$ and
$n$. The local samples also show that local ULIRGs typically have higher $G_0$ than
the mostly LIRG and sub-LIRG star-forming galaxies (although
there is substantial overlap between the two populations). This is in
contrast with the \cii/\aco\ measurements from \citet{Stacey91} that
suggested that $G_0$ can efficiently distinguish between starburst and
spiral galaxy star-formation. The observations of individual local, mostly (sub-)LIRG,
star-forming galaxies \citep[primarily based on \cii\ and \oi\ data;
][]{Malhotra01} extend to higher $n$ and $G_0$ than the GOALS results,
although they broadly follow the same trend.
The PDR density and FUV field strength from our stacked line
measurements of SMGs align with the regions for both local ULIRGs and
local star-forming galaxies (Figure~\ref{fig:pdrcont}). However, the current data are unable to
distinguish between probable merger triggering and secular evolution
of SMGs.

Figure~\ref{fig:pdrcont} also shows that the $n$ values derived
from our fine-structure line spectroscopy are lower
than most archival measurements of individual SMGs (both lensed and unlensed).
It is possible that the apparent difference between our stacking results and
individual archival SMG studies are due to bias in the selection of individually
analysed SMGs or uncertainties in our
analysis, such as if \oi\ self absorption is significantly greater
than the factor of two that we have investigated. However, it could
also be due to systematic uncertainties in the archival analyses,
since most of the archival measurements are inferred from only two
spectral features, CO and \cii, and thus $n$ is primarily constrained
by CO. These existing SMG studies typically use high-$J$ CO
observations, which are converted to \aco\ fluxes using standard
ratios, but in SMGs \aco\ and high-$J$ emission regions are often not
co-located, with higher-$J$ lines tracing warmer, more compact dust
than \aco\ \citep[e.g.,][]{Ivison11, Riechers11b, Spilker15}.  This
means that for SMGs \aco\ fluxes estimated from higher-$J$
measurements may not trace the same region or spatial scales as the
\cii\ and PDR models, which could bias the derived $n$. 
\citet{Bisbas14}, for example, demonstrate a similar effect, showing
that different CO transitions probe to different cloud depths in the
PDRs in NGC\,4030.

\section{Conclusions}
\label{sec:conc}

We have presented \herschel-PACS spectroscopy and photometry of a
sample of 13 gravitationally lensed SMGs at $z=1.03$--3.27, targeting
the \oiv, \feii, \sxiii, \sixii, \oiii, \niii, and \oi\ fine-structure
lines, and the \htsz\ and \htso\ hydrogen rotational lines. We
detected only two lines at $\ge3\sigma$ significance (\oiii\ in
NGP.NA.144, and \oiv\ in HXMM01).

To supplement our data we identified 32 additional SMGs that also have
\herschel-PACS spectroscopy of the targeted lines, and we stacked
these archival spectra with those from our 13 originally targeted
sources. The stacked spectra include eight (for the two hydrogen
lines) to 37 (for \oiii) SMGs, resulting in average spectra of SMGs
with improvements of up to a factor of $\sim6$ in the nominal noise
level. We detected \oi, \sixii, and \niii\ in the stacks, with line strengths relative
to the far-IR continuum of $(0.36\pm0.12)\times10^{-3}$,
  $(0.84\pm0.17)\times10^{-3}$, and $(0.27\pm0.10)\times10^{-3}$, respectively.

Based on the \oiii/\niii\ line ratios we determined that SMGs
are typically  enriched galaxies, with gas-phase metallicities
$\gtrsim Z_{\sun}$. The stacked \sixii/\sxiii\ ratio indicates that
there is some LINER and/or AGN contribution to the IR spectra,
although the absence of strong \oiv\ emission and other AGN tracers
suggests that these are unlikely to dominate the energetics of typical
SMGs. 

The ratio of the \oi\ flux to the far-IR continuum flux in the stacked
data is $(0.36\pm0.12)\times10^{-3}$, significantly lower than the
roughly $2\times10^{-3}$ observed in local sub-LIRGs, but consistent
with local ULIRGs. We used {\sc PDR Toolbox} to model the stacked \oi\
and \sixii\ data and also included average \cii\ measurements from
\citet{Gullberg15} and \citet{George15}. The model indicates that on
average the PDRs in SMGs have gas densities, $n$, of
$10^{1-3}{\rm cm^{-3}}$ and FUV field strengths,
$G_0=10^{2.2-4.5}$. These values are consistent with both measurements
of local ULIRGs and mostly LIRG and sub-LIRG star-forming
galaxies. Additional IR data are required to further constrain the PDR
models and determine whether the star-formation in high-redshift SMGs
is more similar to local sub-LIRGs than local ULIRGs. The derived $n$
is lower than most measurements of individual SMGs from \cii\ and CO
data, which may be due to the previous widespread use of high-$J$ CO
transitions and the uncertainties converting these to \aco\
luminosities.

\acknowledgments
The authors thank Lee Armus, Jeronimo Bernard-Salas, Kristen Coppin, Tanio
D\'iaz-Santos, Jonatan
Selsing, Ian Smail, Mark Swinbank, and Zhi-Yu Zhang for helpful discussions, and an anonymous referee for constructive and clarifying comments.
We are grateful to Javier Graci\'{a}-Carpio, Eckhard Sturm, and the SHINING
collaboration for providing a compilation of spectral observations and
IR emission measurements used for comparison with our sample, and
to Lucia Marchetti, Steve Crawford, Stephen Serjeant and Andrew Baker
for discussion regarding a followup South African Large Telescope
program targeting lens redshifts (Marchetti et al.\ in prep.).

JLW is supported by a European Union COFUND/Durham Junior Research Fellowship under EU grant agreement number 267209, and acknowledges additional support from STFC (ST/L00075X/1).
Support for this work was provided by NASA through an award issued by
JPL/Caltech. AC and JLW acknowledge support from NSF AST-1313319 and NASA
NNX16AF38G.
HD acknowledges financial support from the Spanish Ministry of Economy and Competitiveness (MINECO) under the 2014 Ram\'on y Cajal program MINECO RYC-2014-15686.
LD, SJM and RJI acknowledge support from European Research Council
Advanced Investigator Grant COSMICISM, 321302; SJM and LD are also
supported by the European Research Council Consolidator Grant {\sc
  CosmicDust} (ERC-2014-CoG-647939, PI: H.\,L.\,Gomez).
AV acknowledges support from the Leverhulme Trust through a Research Fellowship.
The Dark Cosmology Centre is funded by the Danish National Research Foundation.
JLW, AC, HD and DR thank the Aspen Center for Physics for hospitality. This work is supported
in part by the NSF under Grant Numbers PHY-1066293 and AST-1055919.
This research was supported by the Munich Institute for Astro- and
Particle Physics (MIAPP) of the DFG cluster of excellence ``Origin and
Structure of the Universe''.

This research has made use of data from the HerMES project
(\url{http://hermes.sussex.ac.uk}). HerMES is a Herschel Key Programme
utilizing Guaranteed Time from the SPIRE instrument team, ESAC
scientists and a mission scientist. The data presented in this paper will be released
through the \hermes\ Database in Marseille, HeDaM
(\url{http://hedam.oamp.fr/HerMES}).
The Herschel-ATLAS (\url{http://www.h-atlas.org})is a project with Herschel, which is an ESA space
observatory with science instruments provided by European-led
Principal Investigator consortia and with important participation from
NASA.
SPIRE has been developed by a consortium of institutes led by Cardiff Univ. (UK) and including: Univ. Lethbridge (Canada); NAOC (China); CEA, LAM (France); IFSI, Univ. Padua (Italy); IAC (Spain); Stockholm Observatory (Sweden); Imperial College London, RAL, UCL-MSSL, UKATC, Univ. Sussex (UK); and Caltech, JPL, NHSC, Univ. Colorado (USA). This development has been supported by national funding agencies: CSA (Canada); NAOC (China); CEA, CNES, CNRS (France); ASI (Italy); MCINN (Spain); SNSB (Sweden); STFC, UKSA (UK); and NASA (USA).
PACS has been developed by a consortium of institutes led by MPE
(Germany) and including UVIE (Austria); KU Leuven, CSL, IMEC
(Belgium); CEA, LAM (France); MPIA (Germany); INAF-IFSI/OAA/OAP/OAT,
LENS, SISSA (Italy); and IAC (Spain). This development has been supported by the funding agencies BMVIT (Austria), ESA-PRODEX (Belgium), CEA/CNES (France), DLR (Germany), ASI/INAF (Italy), and CICYT/MCYT (Spain).
HCSS / HSpot / HIPE is a joint development (are joint developments) by the Herschel Science Ground Segment Consortium, consisting of ESA, the NASA Herschel Science Center, and the HIFI, PACS and SPIRE consortia.
This research has made use of NASA's Astrophysics Data System.

{\it Facilities:}  \facility{Herschel (PACS)}, \facility{Herschel (SPIRE)}

%~~~~~~~~~~~~~~~~~~~~~~~~~~~~~~~~~~~~~~~~~~~~~~~~~~~~~~~~~~~~~~~~~~~~

\bibliographystyle{apj}
\bibliography{bibtex.bib}  

%~~~~~~~~~~~~~~~~~~~~~~~~~~~~~~~~~~~~~~~~~~~~~~~~~~~~~~~~~~~~~~~~~~~~

\appendix 

\section{Archival data included in these analyses}

The archival sources and PACS observations that are included in the stacking analyses are listed in Table~\ref{tab:archsamp}.

\begin{turnpage}
\renewcommand{\tabcolsep}{1mm}
\begin{deluxetable}{lcccccclll}
\tablecaption{Archival sources and observations included in the stacking analyses.
\label{tab:archsamp}} 
\startdata 
\hline\hline  
Name & RA & Dec & $z$ & \lfir$^a$ & \fir$^a$ & Magnification & References$^b$ & Program ID & OBSIDs$^c$ \\ 
 &  &  &  & ($10^{13}$\lsun) & ($10^{-15}{\rm Wm^{-2}}$) & &  &  &  \\ 
\hline
ID9     & $\rm{09^h07^m40\fs0}$ & $-$00\degr42\arcmin01\arcsec & 1.577 & 4.4 & 8.2 & $8.8\pm2.2$  & N14 & OT1\_averma\_1 & 134223228[5--7], 134224524[0--2] \\ 
ID11    & $\rm{09^h10^m43\fs1}$ & $-$00\degr03\arcmin24\arcsec & 1.786 & 5.7 & 7.7 & $10.9\pm1.3$ & N14 & OT1\_averma\_1 & 134223129[1--4]  \\
ID17    & $\rm{09^h03^m03\fs0}$ & $-$01\degr41\arcmin27\arcsec & 2.305 & 6.8 & 5.0 & $4.9\pm0.7$  & N14 & OT1\_averma\_1 & 134223131[3,4] \\
HLock01 & $\rm{10^h57^m51\fs2}$ & +57\degr30\arcmin28\arcsec & 2.958 & 11 & 4.3 & $10.9\pm0.7$ & Co11, R11, S11 & OT1\_averma\_1 & 1342232311, 134224564[4--6], \\
                                                                                                               &&&&&&&&& 1342256251, 1342256261 \\
G15v2.779$^d$ & $\rm{14^h24^m14\fs0}$& +02\degr23\arcmin05\arcsec & 4.243 & 7.6 & 1.2 & $4.1\pm0.2$ & B12, B13 & OT1\_averma\_1 & 1342238160, 1342261470 \\
SMM\,J2135$^e$ & $\rm{21^h35^m11\fs6}$ & +01\degr02\arcmin52\arcsec & 2.326 & 4.3 & 3.6 & $32.5\pm4.5$ & I10 & OT1\_averma\_1 & 1342231704, 1342244443, 1342245235, \\
                                                                                                                     &&&&&&&&& 1342245393, 134225694[0--1], 1342257256 \\
NC.v1.143 & $\rm{12^h56^m32\fs6}$ & +23\degr36\arcmin27\arcsec & 3.565 & 8.9 & 2.2 & $11.3\pm1.7$ & B13, Rp & OT1\_averma\_1 & 1342257[799,800] \\
NA.v1.177 & $\rm{13^h28^m59\fs3}$ & +29\degr23\arcmin27\arcsec & 2.778 & 6.0 & 2.7 & \ldots & B13 & OT1\_averma\_1 & 134225956[0,1] \\
SWIRE\,3-9 & $\rm{10^h43^m43\fs9}$ & +57\degr13\arcmin23\arcsec & 1.735 & 0.47 & 0.73 & 1 & B15 & OT2\_dbrisbin\_1 & 1342253586, 1342253776 \\
SWIRE\,3-14 & $\rm{10^h45^m14\fs5}$ & +57\degr57\arcmin09\arcsec & 1.780 & 0.28 & 0.29 & 1 & B15 & OT2\_dbrisbin\_1 & 1342247014, 1342247131 \\
SWIRE\,4-5 & $\rm{10^h44^m27\fs5}$ & +58\degr43\arcmin10\arcsec & 1.756 & 0.10 & 0.09 & 1 & B15 & OT2\_dbrisbin\_1 & 134224663[8,9]\\
SWIRE\,4-15 & $\rm{10^h46^m56\fs5}$ & +59\degr02\arcmin36\arcsec & 1.854 & 0.30 & 0.37 & 1 & B15 & OT2\_dbrisbin\_1 & 1342253587, 1342253775 \\
SDSS\,J1206 & $\rm{12^h06^m01\fs7}$ & +51\degr42\arcmin28\arcsec & 1.999 & 0.42 & 0.46 & $\sim27$ & B15 & OT2\_dbrisbin\_1 & 1342246801 \\
SMM\,J0302 & $\rm{03^h02^m27\fs7}$ & +00\degr06\arcmin52\arcsec & 1.408 &  0.46 & 1.2 & 1 & B15 & OT2\_dbrisbin\_1 & 134224778[4,5] \\
MIPS\,22530 & $\rm{17^h23^m03\fs3}$ & +59\degr16\arcmin00\arcsec & 1.950 & 0.62 & 0.69 & 1 & B15 & OT2\_dbrisbin\_1 & 1342249495, 1342256260 \\
LESS21  & $\rm{03^h33^m29\fs7}$ & $-$27\degr34\arcmin44\arcsec & 1.235  & 0.07 & 0.13 & 1 & C12 & OT1\_kcoppin\_1 & 1342239701 \\
LESS34  & $\rm{03^h32^m17\fs6}$ & $-$27\degr52\arcmin28\arcsec & 1.098 & 0.06 & 0.14 & 1 & C12 & OT1\_kcoppin\_1 & 1342239703 \\
LESS66  & $\rm{03^h33^m31\fs9}$ & $-$27\degr54\arcmin10\arcsec & 1.315  & 0.14  & 0.41 & 1 & C12 & OT1\_kcoppin\_1 & 1342239369 \\
LESS88  & $\rm{03^h31^m54\fs8}$ & $-$27\degr53\arcmin41\arcsec & 1.269  & 0.08 & 0.17 & 1 & C12 & OT1\_kcoppin\_1 & 1342239705 \\
LESS106 & $\rm{03^h31^m40\fs2}$ & $-$27\degr56\arcmin22\arcsec & 1.617  & 0.15  & 0.23 & 1 & C12 & OT1\_kcoppin\_1 & 1342239753 \\
LESS114 & $\rm{03^h31^m51\fs1}$ & $-$27\degr44\arcmin36\arcsec & 1.606  & 0.33  & 0.57 & 1 & C12 & OT1\_kcoppin\_1 & 1342239702 \\
SPT\,0538-50 & $\rm{05^h38^m16\fs5}$ & $-$50\degr30\arcmin50\arcsec & 2.782 & 5.1 & 2.3 &  $21.0\pm4.0$ & Bo13 & OT2\_dmarrone\_2 & 1342270691 \\          
SPT\,0125-47 & $\rm{01^h25^m07\fs0}$ & $-$47\degr23\arcmin57\arcsec & 2.515 & 8.7 & 4.9 & $5.5\pm0.1$ & A16, V13, We13 & OT2\_dmarrone\_2  & 1342270768 \\ 
SPT\,0103-45 & $\rm{01^h03^m11\fs4}$ & $-$45\degr38\arcmin54\arcsec & 3.092 & 3.4 & 1.2 & $7.2\pm5.2$ & G15, V13, We13, S16 & OT2\_dmarrone\_2 &1342271050 \\  % magnification spilker+inprep:  $7.2\pm2.1$ (avg 2 comps)
F10214 & $\rm{10^h24^m34\fs6}$ & +47\degr09\arcmin09\arcsec &  2.286 & 8.4 & 6.0 & $\sim12$ & S10 & SDP\_kmeisenh\_3  & 1342186812, 1342187021\\
SMM\,J14011 & $\rm{14^h01^m04\fs9}$ & +02\degr52\arcmin24\arcsec & 2.565 & 1.4 & 0.76 & $3.5\pm0.5$ & S05, S13 & KPGT\_kmeisenh\_1 & 134221331[1--4], 1342213677 \\
SMM\,J22471 & $\rm{22^h47^m10\fs4}$ & $-$02\degr05\arcmin53\arcsec & 1.158 & 2.7 & 10 & $\sim2$ & S10 & OT1\_gstacey\_3 & 1342211842, 1342212211 \\
SWIRE\,J104738 & $\rm{10^h47^m38\fs3}$ & +59\degr10\arcmin10\arcsec & 1.958 & 0.40 & 0.42 & 1 & S10 & OT1\_gstacey\_3 & 134223226[8,9] \\
SWIRE\,J104704 & $\rm{10^h47^m05\fs1}$ & +59\degr23\arcmin33\arcsec & 1.954 & 1.0 & 1.1 & 1 & S10 & OT1\_gstacey\_3 & 134223227[0,1] \\
SMM\,J123634 & $\rm{12^h36^m34\fs6}$ & +62\degr12\arcmin41\arcsec & 1.222 & 0.54 & 1.8 & 1 & S10 & OT1\_gstacey\_3 & 1342232[599,601] \\
MIPS\,J142824 & $\rm{14^h28^m24\fs1}$ & +35\degr26\arcmin18\arcsec & 1.325 &  1.3 & 3.6 & $<10$ & HD10, S10 & SDP\_esturm\_3 &  1342187779 \\
SMM\,J02396 & $\rm{02^h39^m56\fs7}$ & $-$01\degr34\arcmin24\arcsec &  1.062 & 0.17 & 0.84 & $\sim2.3$ & G05, C11, C13 & KPGT\_esturm\_1 & 1342214674 
\enddata 
\tablecomments{
$^a$ \lfir\ and \fir\ are the apparent (i.e.\ not corrected for
lensing) far-IR luminosity (40--500\,\micron) and continuum flux
(42.5--122.5\,\micron), respectively (Section~\ref{sec:seds}). Where unavailable from published SED fits these are estimated by scaling published infrared luminosities to the required rest-frame wavelength ranges using the average fitted SED of the targeted SMGs. 
$^b$ References are not the complete list of published papers on each source, but rather the source of data used in this paper:
A16: \citet{Aravena16}, Bo13: \citet{Bothwell13}, B12:
\citet{Bussmann12}, B13: \citet{Bussmann13}, B15: \citet{Brisbin15},
C11: \citet{Chen11}, Co11: \citet{Conley11}, C12: \citet{Coppin12},
C13: \citet{Chen13}, G05: \citet{Greve05}, G15: \citet{Gullberg15},
HD10: \citet{HaileyDunsheath10}, I10: \citet{Ivison10b}, N14:
\citet{Negrello14}, R11: \citet{Riechers11}, Rp: Riechers et al.\ (in
prep.), S05: \citet{Smail05}, S10: \citet{Stacey10}, S11:
\citet{Scott11}, S13: \citet{Sharon13}, S16: \citet{Spilker16}, V13: \citet{Vieira13}, We13: \citet{Weiss13}.
$^c$ OBSIDs are the \herschel\ observation identification number(s) for this program, used to identify the the photometric and spectroscopic observation of each target in the \herschel\ archive.
$^d$ Also known as ID15.141 or ID141.  }
\end{deluxetable}
\end{turnpage}

\section{1D spectra}
\label{app:lines} 

\begin{figure*}[h]
\includegraphics[width=18cm]{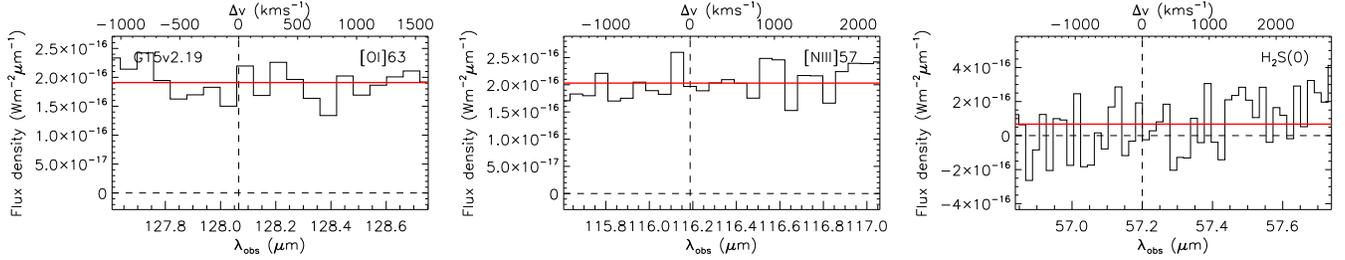}
\caption{{\bf1} 
Observed spectra (black) at native PACS resolution, and the
  best-fit continuum or $\ge3\sigma$ significance Gaussian line
  profiles (solid red) of all targeted
  emission lines for G15v2.19.
  The fitting procedure is detailed in
  Section~\ref{sec:specmeasure}. Broken Gaussian profiles highlight
  potential 1--3$\sigma$ line emission, with the
  significance of each fit reported in brackets (red) in the relevant
  panels. Although this low SNR emission is not discussed in the
  paper, the fits are included here to demonstrate our fitting routine
  and significance of any potential line emission. 
 Dashed black horizontal and vertical lines represent
  the zero continuum level and the expected wavelength of the line based
  on the nominal redshifts from Table~\ref{tab:targs}.}
  \label{fig:lines}
\end{figure*}

In Figure~\ref{fig:lines} (fig~\ref{fig:lines}.1--\ref{fig:lines}.13)
we show the individual spectral observations for each lensed SMG
observed in OT2\_jwardlow\_1. 

\begin{figure*}[h]
\figurenum{\ref{fig:lines}.2}
\includegraphics[width=18cm]{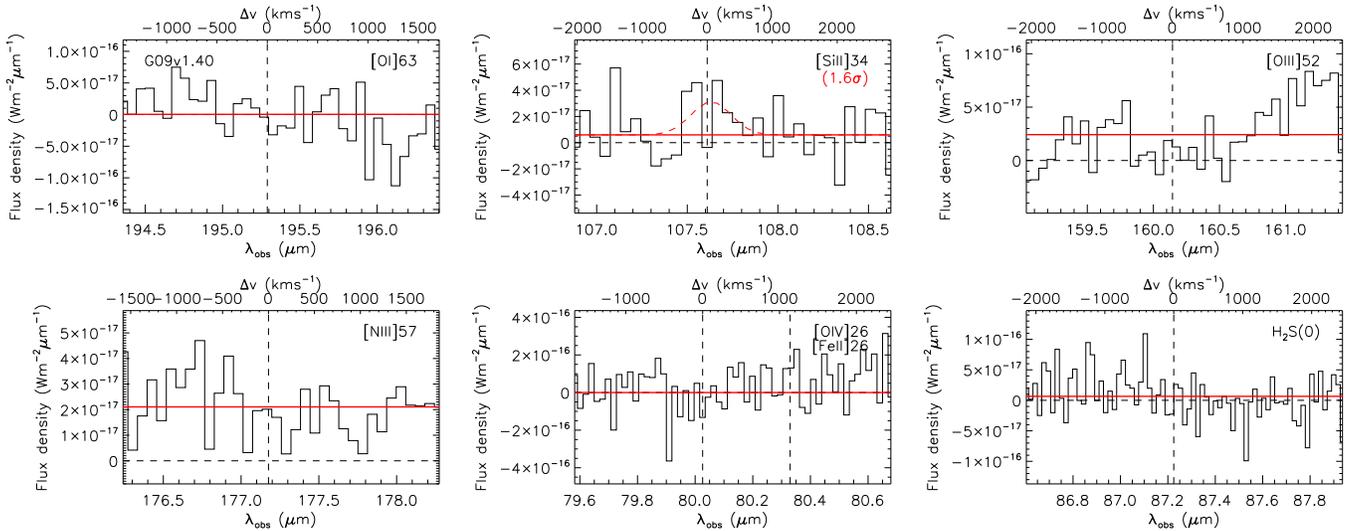} 
\caption{As Figure~\ref{fig:lines}.1 for G09v1.40.}
\end{figure*}
\begin{figure*}[h]
\figurenum{\ref{fig:lines}.3}
\includegraphics[width=18cm]{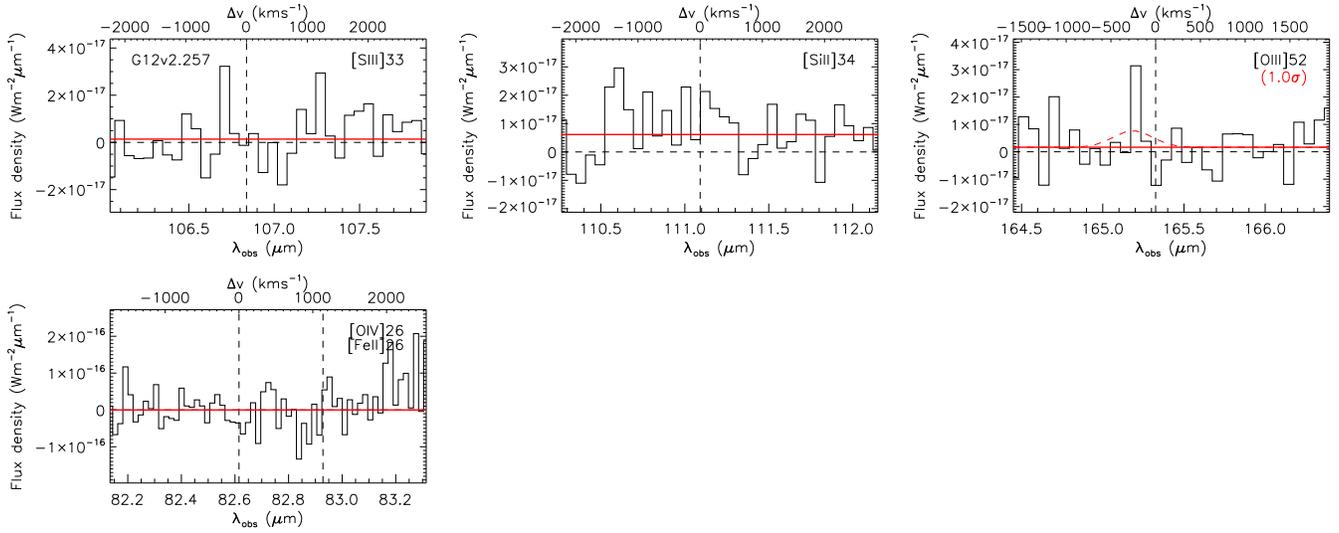}
\caption{As Figure~\ref{fig:lines}.1 for G12v2.257.}
\end{figure*}
\begin{figure*}[h]
\figurenum{\ref{fig:lines}.4}
\includegraphics[width=18cm]{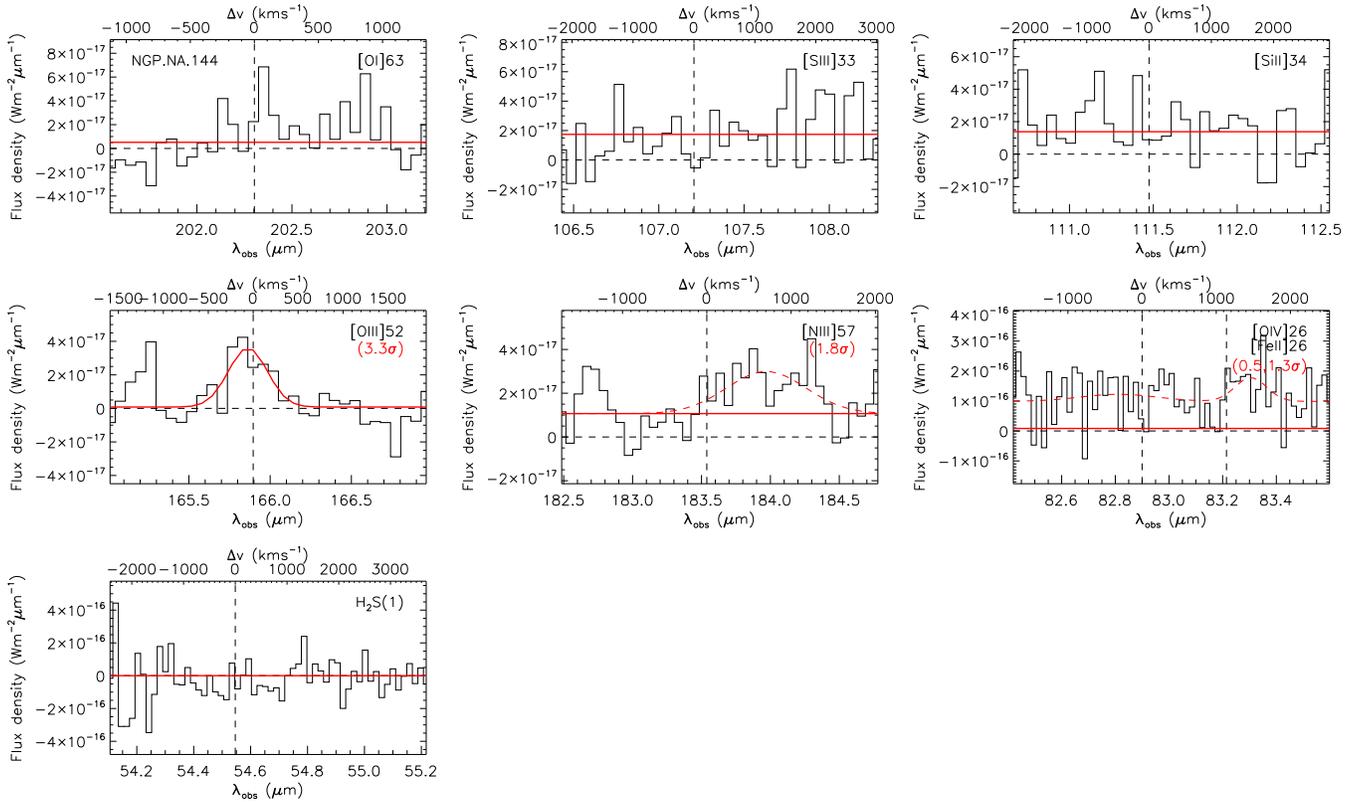}
\caption{As Figure~\ref{fig:lines}.1 for NGP.NA.144.}
\end{figure*}
\begin{figure*}[h]
\figurenum{\ref{fig:lines}.5}
\includegraphics[width=18cm]{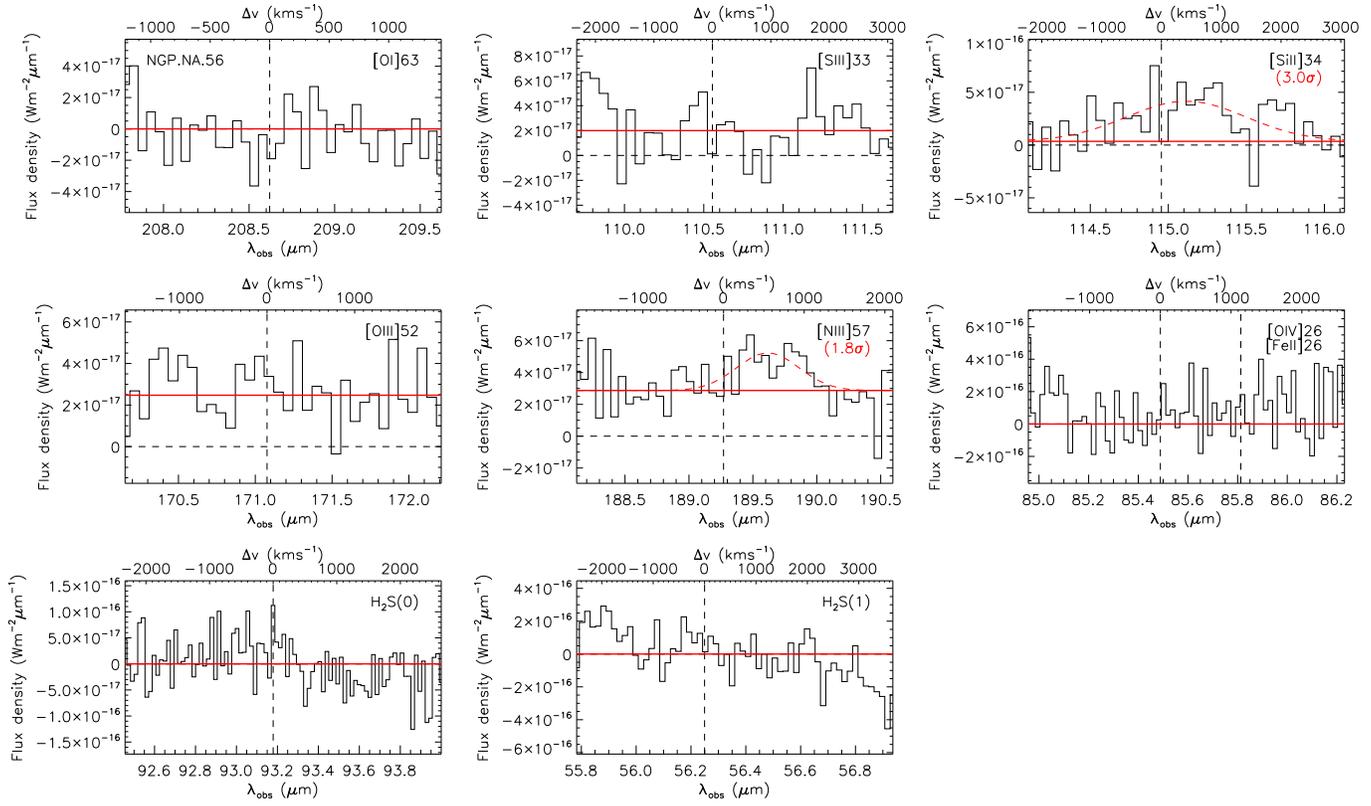}
\caption{As Figure~\ref{fig:lines}.1 for NGP.NA.56.}
\end{figure*}
\begin{figure*}[h]
\figurenum{\ref{fig:lines}.6}
\includegraphics[width=18cm]{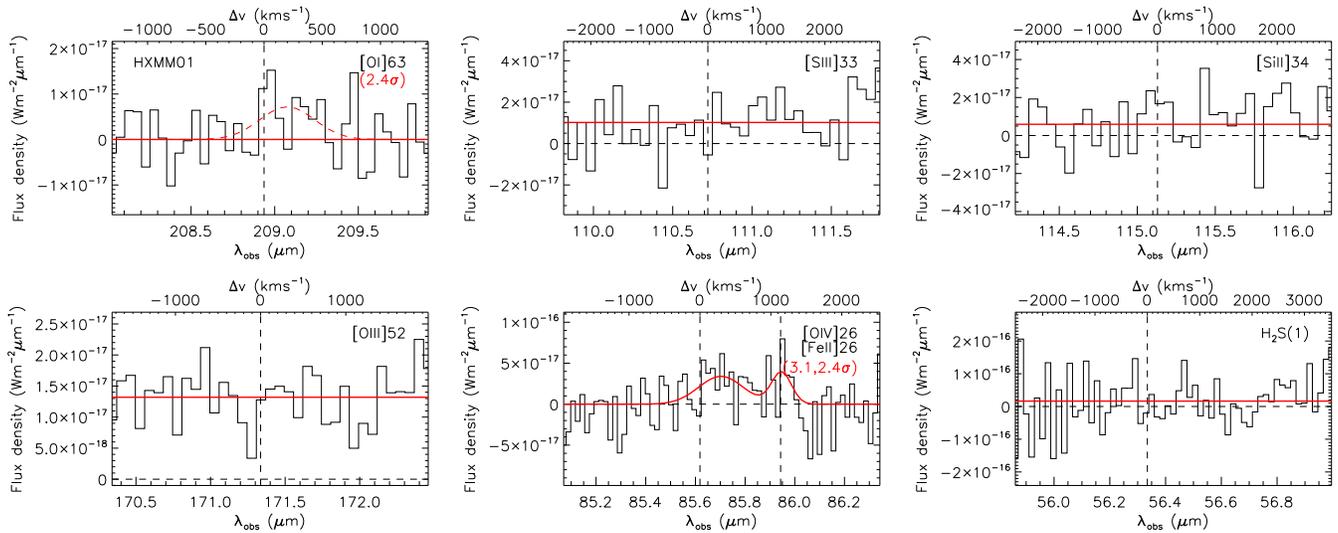}
\caption{As Figure~\ref{fig:lines}.1 for HXMM01.}
\end{figure*}
\begin{figure*}[h]
\figurenum{\ref{fig:lines}.7}
\includegraphics[width=18cm]{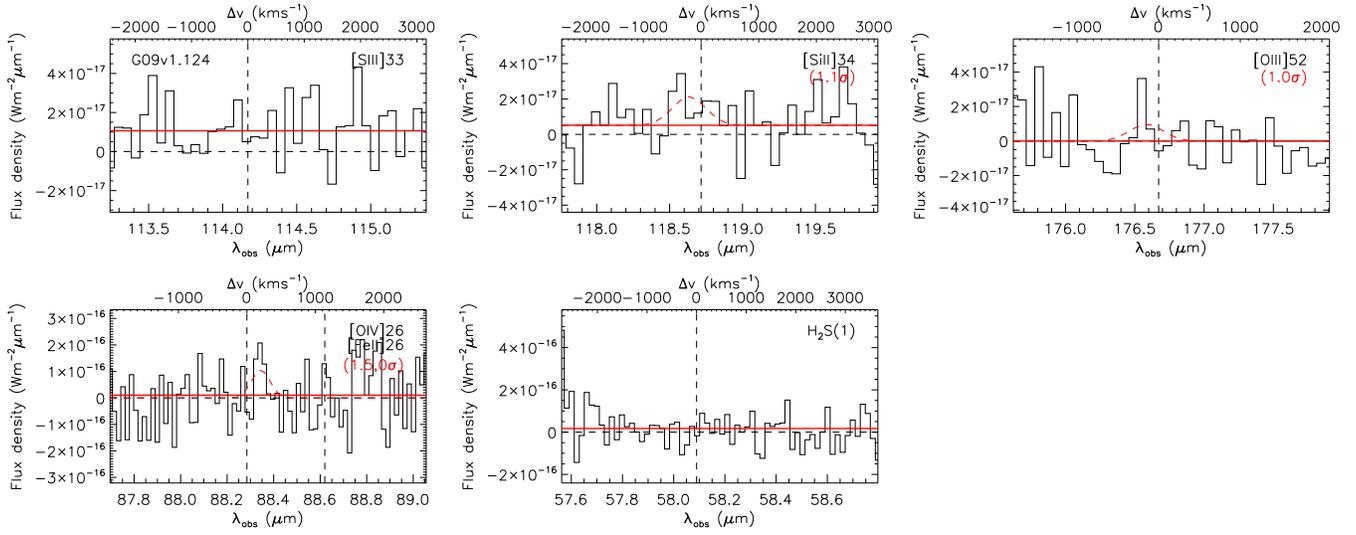}
\caption{As Figure~\ref{fig:lines}.1 for G09v1.124.}
\end{figure*}
\begin{figure*}[h]
\figurenum{\ref{fig:lines}.8}
\includegraphics[width=18cm]{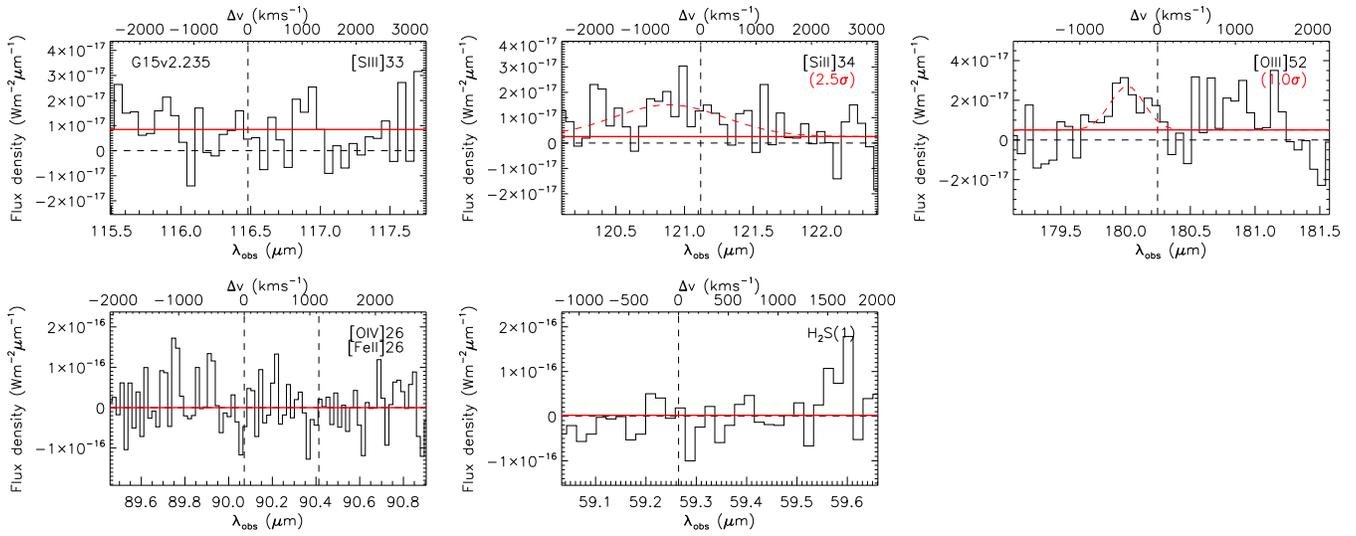}
\caption{As Figure~\ref{fig:lines}.1 for G15v2.235.}
\end{figure*}
\begin{figure*}[h]
\figurenum{\ref{fig:lines}.9}
\includegraphics[width=18cm]{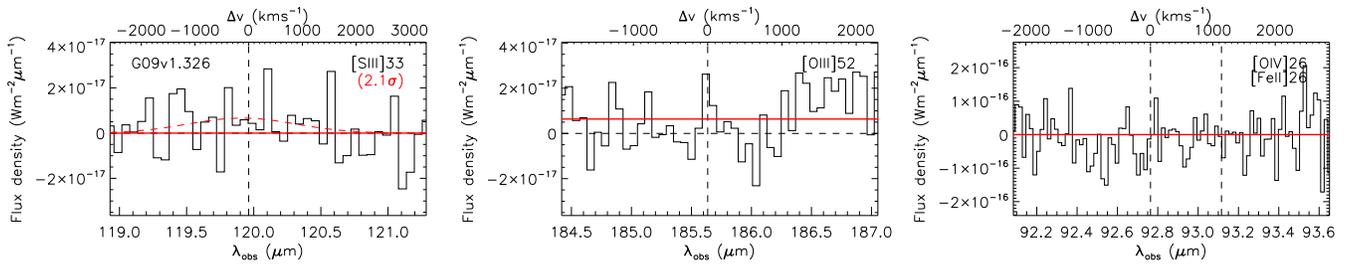}
\caption{As Figure~\ref{fig:lines}.1 for G09v1.326.}
\end{figure*}
\begin{figure*}[h]
\figurenum{\ref{fig:lines}.10}
\includegraphics[width=18cm]{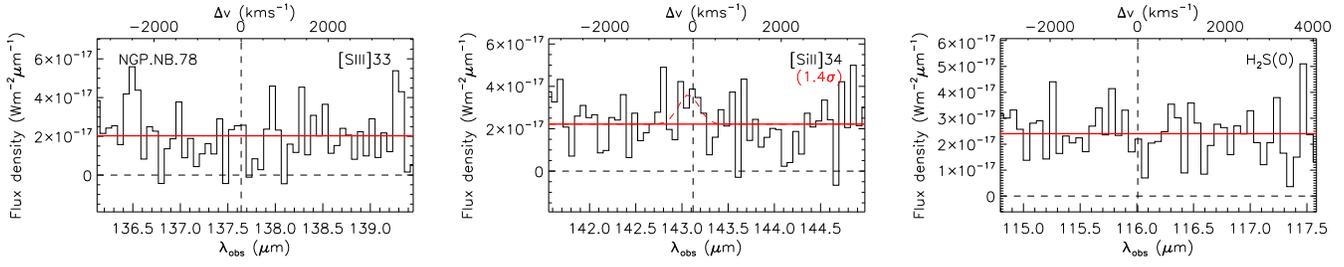}
\caption{As Figure~\ref{fig:lines}.1 for NGP.NB.78.}
\end{figure*}
\begin{figure*}[h]
\figurenum{\ref{fig:lines}.11}
\includegraphics[width=18cm]{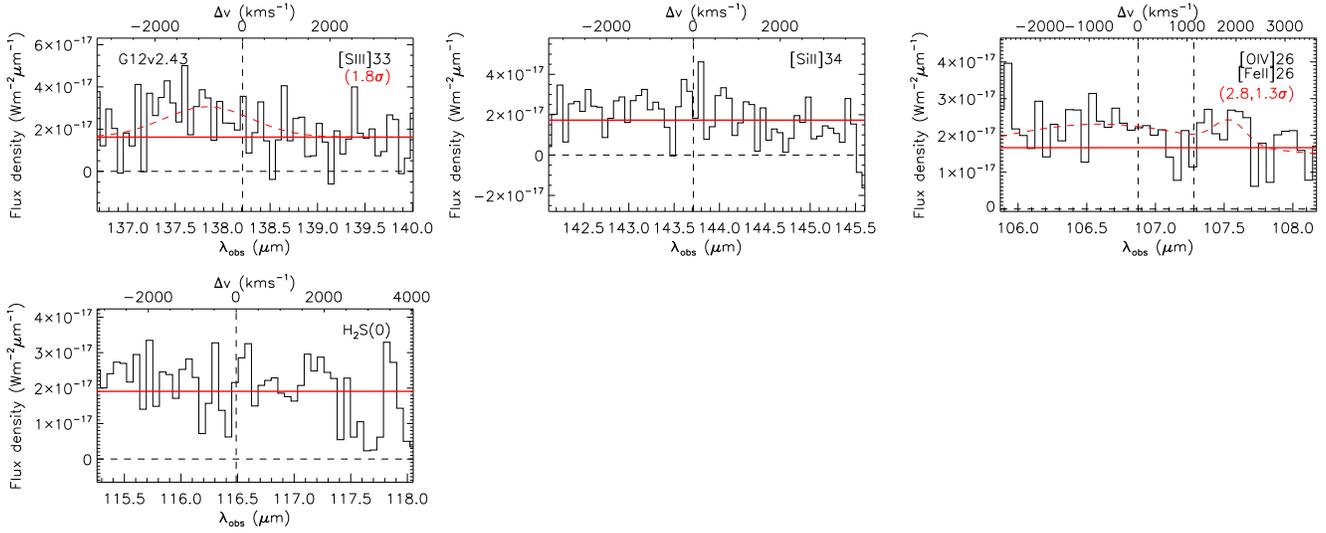}
\caption{As Figure~\ref{fig:lines}.1 for G12v2.43.}
\end{figure*}
\begin{figure*}[h]
\figurenum{\ref{fig:lines}.12}
\includegraphics[width=18cm]{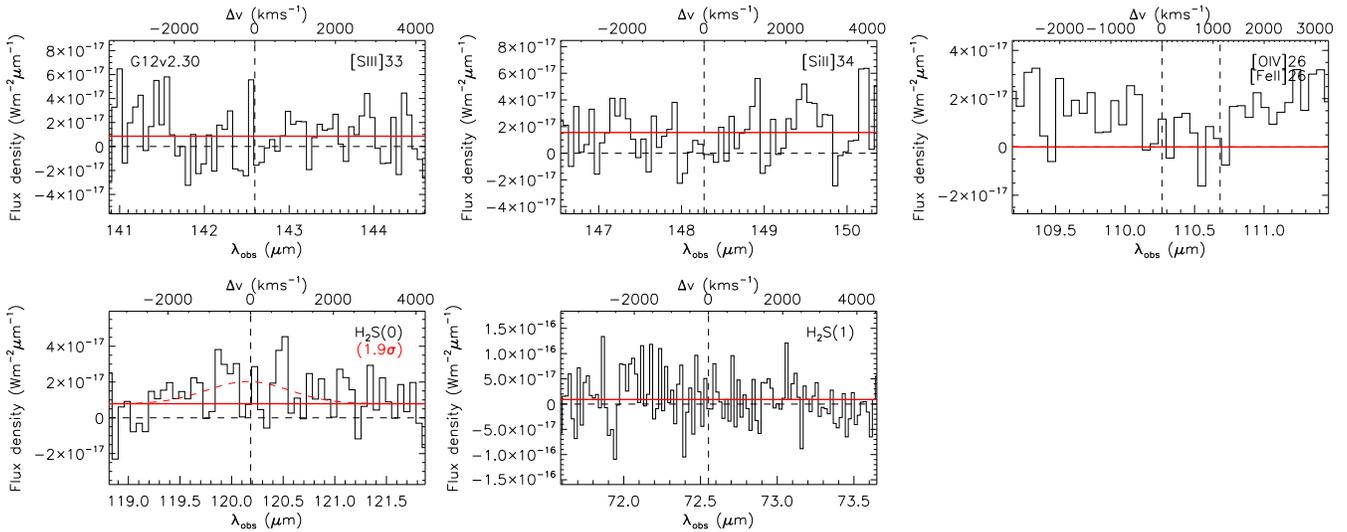}
\caption{As Figure~\ref{fig:lines}.1 for G12v2.30.}
\end{figure*}
\begin{figure*}[h]
\figurenum{\ref{fig:lines}.13}
\includegraphics[width=18cm]{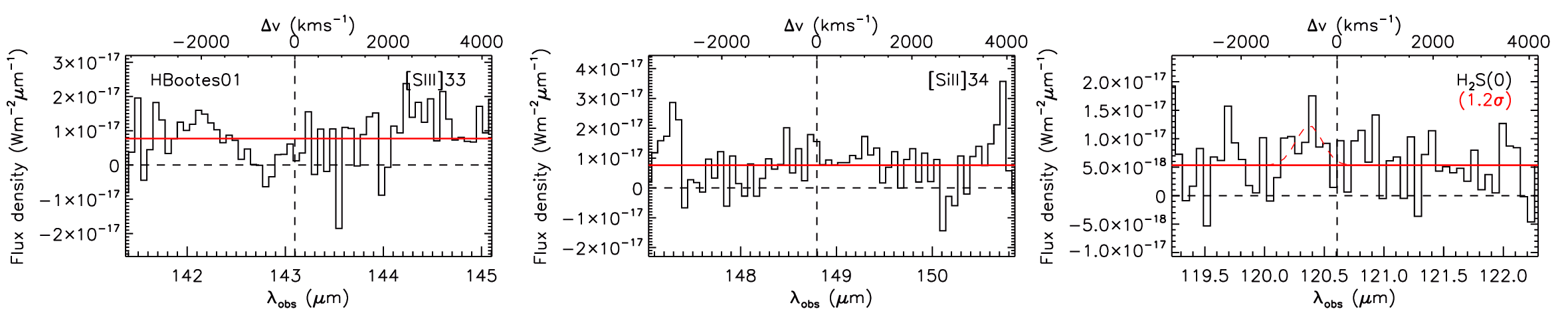}
\caption{As Figure~\ref{fig:lines}.1 for H\bootes01.}
\end{figure*}

%~~~~~~~~~~~~~~~~~~~~~~~~~~~~~~~~~~~~~~~~~~~~~~~~~~~~~~~~~~~~~~~~~~~~

\end{document}